\documentclass [12 pt] {article}
\usepackage {amsmath}
\usepackage{amsthm, amssymb}
\usepackage {graphicx}
\usepackage{url}
\usepackage{fullpage}
\usepackage{multicol}
\usepackage{hyperref}
\usepackage{subfig}
\usepackage{xcolor}

\newtheorem{thm}{Theorem}[section]

\newtheorem{rem}{Remark}[section]

\def\ds{\displaystyle}

	\def\Bn{{\bf n}}

	\def\Bx{{\bf x}}
	\def\By{{\bf y}}
	\def\Bz{{\bf z}}

	\def\B0{{\bf 0}}
	\def\calK{{\mathcal K}}
	\def\calP{{\mathcal P}}
	\def \calD{{\mathcal D}}
	\newcommand{\GD}{\Delta}

	\def \RR {{\mathbb R}}
	\def \CC {{\mathbb C}}
	
	\def \ba {\begin{array}}
	\def \ea {\end{array}}

\title {\bf \Large Active manipulation of Helmholtz scalar fields: Near field synthesis with directional far field control}
\author{Neil Jerome A. Egarguin$^{1,2}$, Daniel Onofrei$^1$, Chaoxian Qi$^3$, Jiefu Chen$^3$ \\ \small{$^1$Department of Mathematics, University of Houston, Houston, TX, USA} \\ \small{$^2$Institute of Mathematical Sciences and Physics, University of the Philippines Los Ba\~nos}\\ \small{College, Laguna, Philippines}\\ \small{$^3$Department of Electrical and Computer Engineering, University of Houston, Houston, TX, USA} }

\date{}

\begin {document}

\maketitle

\begin{abstract}
In this article, we propose a strategy for the active manipulation of scalar Helmholtz fields in bounded near field regions of an active source while maintaining desired radiation patterns in prescribed far field directions. This control problem is considered in two environments: free space and respectively, homogeneous ocean of constant depth. In both media, we proved the existence of and characterized the surface input, modeled as Neumann data (normal velocity) or Dirichlet data (surface pressure) such that the radiated field satisfies the control constraints. We also provide a numerical strategy to construct this predicted surface input by using a method of moments-approach with a Morozov discrepancy principle-based Tikhonov regularization. Several numerical simulations are presented to demonstrate the proposed scheme in scenarios relevant to practical applications.
\end{abstract}

\section{Introduction }
\label{intro}

The active control of acoustic fields in various media has been a very active area of research due to the multitude of possible practical applications. These include the creation of personal audio systems or multizone sound synthesis and reproduction (\cite{Omoto2015, Mahesh2019, Zhang2019, Galvez2019} and references therein), acoustic imaging (\cite{Emerson2013, Laureti2016, Fadden2018} and references therein), active noise control (\cite{Kim2017, Mao2018, Chen2012, Kajikawa2012, Kajikawa2014} and references therein) and acoustic shielding and cloaking (\cite{Onofrei2014, Majid2018, Eggler2019, Eggler2019WM, JCheer2019, Cheer2016} and references therein). In particular, the manipulation of Helmholtz fields in underwater environments is widely-studied as it presents important applications such as in communications, ocean imaging and remote sensing, marine ecosystem monitoring \cite{Zora2011} and military and defense applications \cite{Kuperman2004} (see also the monograph \cite{Kuperman2011} for a detailed discussion of computational strategies for ocean acoustics). In \cite{Proakis2001, Chitre2008}, comprehensive discussions of the development of underwater acoustic networks and the challenges involved were provided. The complexity of the make-up of the ocean environment requires substantial modification of control strategies designed for free space (or other simple media). As such, existing free space strategies are adapted to simpler marine environments like shallow water or a homogeneous finite-depth ocean (see for example the reference monographs \cite{MarineAcoustics} and \cite{Kuperman2011}). For instance, in \cite{Lin2014}, the authors proposed the use of acoustic contrast control strategies to focus sound in shallow water. In the same environment, the works \cite{Buck1997, Peng2014} developed a single-mode excitation with a feedback control algorithm to achieve both near and far field sound control. 

The manipulation of the acoustic far field poses several challenges, such as loss of evanescent fields and diffraction limits. In \cite{Ma2019}, the authors broached a method to overcome these limits and attained effective far field imaging using wave vector filtering. In \cite{Lerosey2007} far-field time reversal was used to overcome those challenges. Another approach is the smart design of transducers  with adaptive structures such as classical rectangular panels in an infinite baffle \cite{Pan1992}, foldable tessellated star transducers \cite{Zou2017}, Helmholtz resonators with computerized controls \cite{Lemoult2011} and modern metamaterials \cite{Nejad2019}.

From the numerical point of view, finite-element methods (FEM) have been continually refined to address some of the shortcomings of the classical FEM, such as those encountered involving acoustic scattering in unbounded domains, numerical dispersion errors and heavy computational requirements especially for adaptive methods. Some recent advances on this front can be found in \cite{Thompson2006} and \cite{Hervellanieto2019}. Several numerical methods employing optimization frameworks are also used especially in solving acoustic inverse problems for biomedical imaging \cite{Poudel2019}, subsurface imaging \cite{Colton2000Review} and sound propagation in waveguides \cite{Kirby2008}. Other approaches include wave-domain methods (as used in \cite{Han2018} and \cite{Han2019}) and modal-domain approaches (for instance \cite{Menzies2012} and \cite{Zhang2016}). The approach employed in this paper (as well as previous works such as  \cite{Onofrei2014}, \cite{Platt2018}, \cite{Hubenthal2016}, \cite{Egarguin2018} and \cite{EgarguinWM2020}) is the use of the Green's function to represent the solution to the Helmholtz equation in terms of a propagator operator and then employ a Tikhonov regulariation scheme with the Morozov discrepancy principle to solve the resulting operatorial equation. For underwater acoustic control problems, three strategies are commonly used in expressing the propagated field (see \cite{Kuperman2011} and \cite{Keller1997} for a discussion of each approach), namely normal modes, the Hankel transform and the ray representation (or the multiple reflection representation for stratified oceans).

In our previous works \cite{Onofrei2014, Platt2018, Egarguin2018}, we used global basis representation of the desired inputs in the spirit of \cite{Doicu}. In this paper, we present new theoretical results and propose novel numerical schemes on the active control of acoustic fields in free space and in a homogeneous ocean of finite depth. These results enhance our previous works \cite{Onofrei2014}, \cite{Platt2018} and \cite{Egarguin2018} by allowing for additional constraints on the fields' radiated far field pattern. Thus, in this work we are able to prove and numerically validate the possibility of characterizing active sources (represented as surface pressure or normal velocities) so that the field generated will approximate given patterns in prescribed exterior regions while maintaining desired far field radiation in fixed directions. 

The main novelty of this paper is the \underline{simultaneous} active control of near fields in prescribed exterior bounded regions and various far field directions with different prescribed far field patterns in two separate environments: free space and homogeneous finite-depth ocean environment. In \cite{Onofrei2014}, \cite{Platt2018} and \cite{Egarguin2018}, only near region field control and the case of an almost nonradiating source were considered.  In the latter, a null field was prescribed in the entire far field region which is a far stronger condition than the one considered in the current study where we allow different far field radiation patterns to be prescribed in different fixed far field directions. This additional constraint gave rise to a new functional framework and additional layers in the numerical scheme. Moreover, \cite{Platt2018} only offered a brief discussion of the theoretical results for the active acoustic control in homogeneous finite-depth ocean environment and did not provide any numerical investigations. Last, but not least we propose here the use of local basis functions to represent the unknown boundary input instead of global basis functions (e.g., spherical harmonics) as used in the aforementioned works. This improved the computation time required for the simulations, especially since the additional far field constraints significantly increased the problem's complexity. This choice may also aid in the physical instantiation of the calculated boundary input as fewer degrees of freedom are now needed to achieve good control accuracy.

The rest of the paper is organized as follows. Section \ref{problem} formally states the mathematical formulation of the general problem. Section \ref{freespace} and Section \ref{oceanenv} present the analysis and numerical results in free space and homogeneous finite-depth ocean environments, respectively. Both of these sections includes subsections discussing the theoretical results and numerical simulations. We end with concluding statements and future research directions in Section \ref{conclusions}.

\section{Statement of the problem}
\label{problem}

We consider the problem of characterizing an active source (modeled as surface pressure or surface normal velocity) to accurately approximate a priori given fields in several bounded exterior regions while synthesizing different desired patterns in various prescribed far field directions. Let $D_a\Subset R$ (where $R$ denotes the environment space to be defined below and $\Subset$ denotes a compact embedding) be the active source modeled as a compact region in space with Lipschitz continuous boundary and $\{R_1, R_2,...,R_m\}$ be a collection of $m$ mutually disjoint smooth domains exterior to $D_a$. Moreover, we consider $n$ distinct directions $\hat \Bx_1, \hat \Bx_2,..., \hat \Bx_n$ representing the far field directions of interest. Mathematically, the problem is to find a boundary input on the source, either a Neumann input data $v \in C(\partial D_a )$ (normal velocity) or a Dirichlet data $p \in C(\partial D_a)$ (pressure) such that for any desired field $f = (f_1, f_2,..., f_m)$ on the control regions (i.e., for each $l$, $f_l$ solves the homogeneous Helmholtz equations in some neighborhood of $R_l$) and prescribed far field pattern values $f_\infty = (f_{\infty,1}, f_{\infty,2},..., f_{\infty,n})$, the solution $u$ of the following exterior Helmholtz problem:
\begin{equation}
\label{P1a}
\vspace{0.15cm}\left\{\vspace{0.15cm}\begin{array}{llll}
\GD u+k^2u=0 \mbox{ in } R \!\setminus D_a \vspace{0.15cm},\\
\nabla u\cdot \Bn=v, (\mbox{ or } u=p)\mbox{ on }\partial D_a\\
\text{boundary conditions corresponding to the medium}\\
\text{suitable radiation condition in the medium}
 \end{array}\right.
\end{equation}
and its corresponding far field pattern $u_\infty$ satisfies 
\begin{equation} \label{P1b} 
\begin{cases}
\Vert u-f_l\Vert_{C^2(R_l)}\leq \mu \text{ for } l= \overline{1,m} \\
|u_\infty(\hat \Bx_j)-f_{\infty,j}| \leq \mu \text{ for } j= \overline{1,n}
\end{cases}
 \end{equation}  
for a desired small positive accuracy threshold $\mu$. Here and throughout the rest of the paper $\Bn$ is the outward unit normal to $\partial D_a$ and ${\hat{\Bx}}=\frac{\Bx}{|\Bx|}$ denotes the unit vector along the direction $\Bx$. Moreover, the $e^{-i \omega t}$-dependence of the fields, where $\omega = kc$ and $c$ is the propagation speed of sound in the respective media, is implicitly assumed and omitted. For the free space environment, $R =\RR^3$, the radiation condition is 
\begin{equation} \label{radiation-fs}\displaystyle\left<{\hat{\Bx}},\nabla u(\Bx)\right>\! -\!iku(\Bx)\!=\!o\left(\frac{1}{|\Bx|}\!\right)\!,\mbox{ as }|\Bx|\rightarrow\infty
\mbox{ uniformly for all ${\hat{\Bx}}$ } \end{equation} and there are no additional boundary conditions. Meanwhile the underwater environment is modeled as an homogeneous ocean with constant depth $-h>0$ (see \cite{MarineAcoustics}) and we have $R= \{\Bx = (x, y,z) \in \RR^3~|~ z \in [h, 0]\}$ with medium boundary conditions
\begin{equation}
\label{bc-ocean0}
\begin{cases}
u &=0 \text{ at the ocean surface } z=0 \text{ and } \\
 \dfrac{\partial u}{\partial z} &= 0 \text{ at the ocean floor }z=h
 \end{cases}.
 \end{equation} The radiation condition for this environment is given in Section \ref{ocean_theory}. 

Classical results (for instance, \cite{ColtonKress2013}, \cite{MarineAcoustics}) guarantee that for every set of given Dirichlet or Neumann inputs on $\partial D_a$, problem \eqref{P1a} has a unique continuous radiating solution $u$ (with the additional condition that the normal derivative exists in the sense of uniform convergence for the Neumann problem). Building-up the strategy used in \cite{Onofrei2014, Platt2018, Egarguin2018} we analyze a representation for the unique solution of the above exterior problem as a function of the inputs and use this to characterize the boundary data that will ensure \eqref{P1b}. We consider a fictitious source $D'_a \Subset D_a$ and slightly larger mutually disjoint open regions $W_1, W_2,..., W_m$ with $R_l \Subset W_l$. We assume that for each $l$, $f_l$ solves the homogeneous Helmholtz equations in $W_l$ and also assume that the larger regions and the source are well separated, i.e., \begin{equation} \label{separation} \overline{W_l} \cap \overline{D_a} = \emptyset, \text{ for } l=\overline{1,m}. \end{equation} Lastly, we let $Y= \displaystyle \prod_{l=1}^m L^2(\partial W_l)$ be the $L^2$ space of $m$-tuples of functions on the $W_l$'s with the inner product \begin{equation} <\varphi, \psi>_Y = \sum_{l=1}^m <\varphi_l, \psi_l>_{L^2(\partial W_l)}
\end{equation}
 for all $\varphi = (\varphi_1, \varphi_2,..., \varphi_m)$ and $\psi=(\psi_1, \psi_2,... , \psi_m) \in Y$.

In the next two sections, we shall present the theoretical formulation and proof of the existence with explicit characterization of a class of solutions to the above questions backed with numerical simulations showing the feasibility of such a control scheme, for both the free space and homogeneous finite-depth ocean environment.

\section{Free space environment}
\label{freespace}

In this section, we shall deal with the problem of controlling the near field in various bounded exterior regions of space while creating prescribed far field patterns in several directions in a free-space environment using a single active source. We begin with the establishment of a proof of the existence of a solution for the active control problem \eqref{P1a}-\eqref{P1b}, \eqref{radiation-fs}. Then we propose a strategy for its explicit characterization and building on the numerical scheme developed in \cite{Platt2018, Egarguin2018, EgarguinWM2020} we produce simulations supporting the current theoretical results.

\subsection{Theoretical Framework}
It was shown in \cite{Onofrei2014}, \cite{Platt2018} that  if $k$ is not a resonance wavenumber (see \cite{colton_kress} and \cite{Onofrei2014}, \cite{Platt2018}), the normal velocity $v$ or pressure $p$ on the surface $\partial D_a$ of the active source needed to solve  the control problem  \eqref{P1a}, \eqref{P1b}, \eqref{radiation-fs} can be characterized by a density $w \in L^2(\partial D'_a)$ such that
\begin{eqnarray}
& v(\By) =&\displaystyle \frac{-i}{\rho c k}\frac{\partial}{\partial\Bn}\int_{\partial D'_{a}}w(\Bx) \phi(\Bx, \By) dS_\Bx \text{ and }\label{eqnvn}\\
\nonumber \\ 
& p(\By)=&\displaystyle \int_{\partial D'_{a}}w(\Bx) \phi(\Bx,\By) dS_\Bx,\label{eqnpb}
\end{eqnarray} 
where $\rho$ denotes the density of the surrounding environment, $c$ denotes the speed of sound in the given media and $\phi(\Bx, \By) = \dfrac{e^{ik|\Bx-\By|}}{4 \pi |\Bx-\By|}$ is the fundamental solution of the 3D Helmholtz equation. The motivations behind \eqref{eqnvn} and \eqref{eqnpb} are summarized in the following remarks.

\begin{rem} The expressions in \eqref{eqnvn} and \eqref{eqnpb} provide an ansatz for solutions of \eqref{P1a}, \eqref{radiation-fs}. This ansatz is then used in a control argument to find a density $w$ on a fictitious source $D_a'$ so that the control constraints in \eqref{P1b} are satisfied.
\end{rem}

\begin{rem}
The use of the fictitious source in the ansatz in \eqref{eqnvn} and \eqref{eqnpb} simplifies the analysis and calculations as $D_a'$ can be chosen to be a sphere compactly embedded in the physical source. Recall that the physical source can assume any compact shape as long as it is well-separated from the control regions and has a Lipschitz continuous boundary to ensure the well-posedness of the exterior Helmholtz problem.
\end{rem}

\begin{rem}
The boundary input obtained from the ansatz  in \eqref{eqnvn} and \eqref{eqnpb} will be smooth. From a theoretical standpoint, this is desirable when the present scalar control results are extended to a vector Helmholtz or a Maxwell system (see \cite{OEP2020}). From an applied perspective, smooth boundary inputs are often more suitable for practical applications as they are easier to approximate.
\end{rem}

Although the expressions in \eqref{eqnvn} and \eqref{eqnpb} make use of the single layer potential operator, it was noted in \cite{Platt2018} (see also \cite{Egarguin2018}) that these inputs can be written in terms of the double layer potential operator and hence, also in terms of  linear combinations of the two. Consequently, the results to be presented can be adapted to the case when the propagator operators are expressed in terms of a linear combination of the single and double layer potentials.

With this density $w  \in L^2(\partial D'_a)$, the field $u$ satisfying \eqref{P1a} can be characterized on each control region by the operator $\calK: L^2(\partial D'_a) \to Y$, with
\begin{equation}
\calK w (\By_1, \By_2,..., \By_m) = \big (\calK_1 w (\By_1), \calK_2 w (\By_2), ..., \calK_m w (\By_m) \big )
\end{equation}
where for each $l =\overline{1,m}$, $\By_l \in \partial W_l$ and 
\begin{equation}
\calK_l w (\By_l) = \displaystyle \int_{\partial D'_{a}}w(\Bx) \phi(\Bx,\By_l) dS_\Bx.
\end{equation}
From \cite{ColtonKress2013}, the solution $u$ has the asymptotic (far field) expression 
\begin{equation}
u(\Bx_0) = \dfrac{e^{ik|\Bx_0|}}{|\Bx_0|} \left ( u_\infty(\hat \Bx_0) + \mathcal O \left (\dfrac{1}{|\Bx_0|} \right ) \right )
\end{equation}
uniformly in the direction $\hat \Bx_0$ as $|\hat \Bx_0| \to \infty$ and where the function $u_\infty$ given by
\begin{equation}
  u_\infty(\hat \Bx_0) = \dfrac{1}{4\pi} \int_{\partial D_a'} w(\By) e^{-i k \mathbf{\hat \Bx_0} \cdot \By} dS_\By
 \end{equation}
is called the \textbf{far field pattern} of $u$. 

\begin{rem}
	\label{int-ctr-free-space}
	The restriction that each $f_l$ satisfies the Helmholtz equation in some neighborhood of $R_l$ and the fact that $R_l \Subset W_l$ for all $1\leq l\leq m$ ensure, through uniqueness and regularity results for the interior Helmholtz problems (in the spirit of \cite{Onofrei2014}), that the field $u$, solution of \eqref{P1a}, \eqref{radiation-fs}, will satisfy the control constraint \eqref{P1b} if 
	$$
	\begin{cases}
	\Vert u-f_l\Vert_{L^2(\partial W_l)}\leq \mu \text{ for } l= \overline{1,m} \\
	|u_\infty(\hat \Bx_j)-f_{\infty,j}| \leq \mu \text{ for } j= \overline{1,n}
	\end{cases}.
	$$
\end{rem}

Hence, from the Remark \ref{int-ctr-free-space} we deduce that the control problem \eqref{P1a}, \eqref{P1b}, \eqref{radiation-fs} amounts to to finding the density $w \in L^2(\partial D'_a)$ so that the corresponding solution $u$ of \eqref{P1a}, \eqref{radiation-fs} and its corresponding far field pattern $u_\infty$ satisfy
\begin{equation}
\begin{cases}
\Vert u - f \Vert_{L^2(\bigcup_{l =1}^m \partial W_l)}\leq \mu \\
|u_\infty(\hat \Bx_j)-f_{\infty,j}| \leq \mu \text{ for } j= \overline{1,n}
\label{farfieldfree}
\end{cases}
\end{equation}
for any  $f = (f_1, f_2,..., f_m) \in Y$ and fixed directions $\hat \Bx_j$, $j =\overline{1,n}$. The second constraint in \eqref{farfieldfree} is an added novelty to our work, as we consider the far field pattern in certain prescribed and a-priori fixed far field directions $(\hat \Bx_1, \hat \Bx_2, ..., \hat \Bx_n)$.  We model the far field pattern in these far field directions by using the far field pattern operator $\calK_{\infty}: L^2(\partial D'_a) \to \CC^n$ defined as \[\calK_\infty w  = \big ( \calP_{w,1}, \calP_{w,2}, ..., \calP_{w,n}\big )\] where for each $j = \overline{1, n}$, 
\begin{equation}
 \calP_{w,j} = \dfrac{1}{4\pi} \int_{\partial D_a'} w(\By) e^{-i k \mathbf{\hat \Bx_j} \cdot \By} dS_\By.
\end{equation}
Hence, the overall propagator operator $\calD: L^2(\partial D'_a) \to Y \times \CC^n$ is defined such that
\begin{equation}
\calD w (\By_1, ..., \By_m) = \big ( \calK_1 w (\By_1), ..., \calK_m w (\By_m),   \calP_{w,1}, \calP_{w,2}, ..., \calP_{w,n} \big ).
\label{propagator_free}
\end{equation}
where $\CC^n$ is endowed with the usual dot product and where $Y \times \CC^n$ is described by the usual graph metric, \[\displaystyle<u,v>_{Y \times \CC^n}=\displaystyle<f,g>_{Y}+\sum_{i=1}^n c_i\cdot \overline{d}_i\] for $u=(f,c_1, c_2,...,c_n), v=(g,d_1, d_2,...,d_n) \in Y \times \CC^n$. 
To show the existence of a solution to the control problem (\ref{P1a}, \eqref{radiation-fs}, \ref{farfieldfree}) we show that the linear compact propagator operator $\calD$ has a dense range. This is established in the following theorem by showing that the adjoint operator $\calD^*$ has a trivial kernel.

\begin{thm}
\label{thm:free}
Except a discrete set of values for $k$, the operator $\calD$ defined in \eqref{propagator_free} has a dense range.
\end{thm}

\begin{proof}
We prove the equivalent assertion that the adjoint $\calD^*$ has a trivial kernel. We first note that by simple algebraic manipulation one can obtain that the adjoint operator $\calD^*: Y \times \CC^n \to L^2(\partial D'_a)$ is given by
\begin{equation}
\big (\calD^*(\psi, c) \big ) (\By) = \sum_{l =1}^m \calK^*_l \psi_l (\By) + \sum_{j=1}^n \dfrac{c_j e^{ik \hat \Bx_j \cdot \By}}{4\pi},
\end{equation}
for any $\psi = (\psi_1, \psi_2,...,\psi_m) \in Y$, $c= (c_1, c_2,...,c_n) \in \CC^n$ and $\By \in \partial D'_a$ where the operator $\calK_l^*: L^2(\partial W_l) \to L^2(\partial D'_a)$ is given by 
\begin{equation}
\calK_l^* \psi_l (\By) = \int_{\partial W_l} \psi_l(\Bx) \overline{\phi} (\Bx, \By) dS_\Bx,
\end{equation}
for $l =\overline{1, m}$. Suppose $(\psi, c) \in \ker \calD^*$, i.e., 
\begin{equation}
 \big (\calD^*(\psi, c) \big ) (\By) =  \sum_{l =1}^m \calK^*_l \psi_l (\By) + \sum_{j=1}^n \dfrac{c_j e^{ik \hat \Bx_j \cdot \By}}{4\pi} =0
 \label{ker_assump}
 \end{equation}
 for any $\By \in \partial D'_a$. Define $w(\By) = \ds \sum_{l =1}^m  \int_{\partial W_l} \overline \psi_l(\Bx) \phi (\Bx, \By) dS_\Bx + \ds \sum_{j=1}^n \dfrac{\overline c_j e^{-ik \hat \Bx_j \cdot \By}}{4\pi}$, where the integrals exists as improper integrals on the $\partial W_l$'s. Note that each term in $w$ is a solution of the Helmholtz equation and so together with \eqref{ker_assump}, we have
 \begin{equation}
 \begin{cases}
 \GD w+k^2w=0 \text{ in } D'_a \\
 w =0 \text{ on }\partial D'_a
 \end{cases}.
 \end{equation}
 Proceeding as in \cite{Onofrei2014}, by using analytic continuation, $w=0$ in $D_a'$ and then by the continuity of the single layer potential together with the uniqueness of the interior problem in each of the regions $\{W_l\}_{l=1}^m$ we obtain that $w = 0$ on $\RR^3$. Finally,  classical interior and exterior jump relations for  the single layer potential on the $\partial W_l$'s imply $\psi_l = 0$ on   $\partial W_l, l=\overline{1, m}$. This, when used in \eqref{ker_assump} gives 
 \begin{equation}
 \sum_{j=1}^n c_j e^{ik \hat \Bx_j \cdot \By} =0
 \label{cj_eqn}
 \end{equation}
 for any $\By \in \RR^3$. We seek to show that $c_j =0$ for $j = \overline{1, n}$. Fix a $\hat \By_0 \in \RR^3$ and define $\By_p = p \hat \By_0$ for $p = \overline{0, n-1}$. Plugging-in $\By = \By_p$ in \eqref{cj_eqn} yields the $n \times n$ system
 \begin{align}
 c_1 + c_2 + ... +c_n &=0 \nonumber  \\
 c_1 e^{ik \hat \Bx_1 \cdot \By_1} +  c_2 e^{ik \hat \Bx_2 \cdot \By_1} +...+  c_n e^{ik \hat \Bx_n \cdot \By_1} &=0 \nonumber \\
 & \vdots   \label{system1} \\
  c_1 e^{ik \hat \Bx_1 \cdot \By_{n-1}} +  c_2 e^{ik \hat \Bx_2 \cdot \By_{n-1}} +...+  c_n e^{ik \hat \Bx_n \cdot \By_{n-1}} &=0.  \nonumber
 \end{align}

 Let $\Bz_j = e^{ik \hat \Bx_j \cdot \hat \By _0}$. Then \eqref{system1} can be written as a Vandermonde system
 \begin{equation}
 \left[
\begin{array}{cccc}
 1 & 1 & \cdots  & 1 \\
 z_1 & z_2 & \cdots  & z_n \\
   &   & \vdots  &   \\
 z^{n-1}_1 & z^{n-1}_2 & \cdots  & z^{n-1}_n \\
\end{array}
\right] \left[
\begin{array}{c}
 c_1 \\
 c_2 \\
 \vdots  \\
 c_n \\
\end{array}
\right] = \left[
\begin{array}{c}
 0 \\
0 \\
 \vdots  \\
 0 \\
\end{array}
\right].
\label{vandermonde1}
 \end{equation}
 This system admits a unique solution (the trivial solution $c_j =0$ for all $j=\overline{1,n}$) unless the coefficient matrix has determinant zero. Note that 
 \[ \text{det} \left ( \left[
\begin{array}{cccc}
 1 & 1 & \cdots  & 1 \\
 z_1 & z_2 & \cdots  & z_n \\
   &   & \vdots  &   \\
 z^{n-1}_1 & z^{n-1}_2 & \cdots  & z^{n-1}_n \\
\end{array}
\right] \right ) = \prod_{1 \le j < l \le n} (z_j-z_l),\]
which is zero if and only if there are indices $q_1$ and $q_2$ such that $z_{q_1} = z_{q_2}$, or equivalently, if and only if \begin{equation} (\hat \Bx_{q_1}-\hat \Bx_{q_2}) \cdot \hat \By_0  = \dfrac{2 \pi}{k} M \label{eqn2} \end{equation} for some integer $M$. By the Triangle and the Cauchy-Schwarz inequalities we obtain \begin{equation}|M| = \dfrac{k}{2 \pi} |(\hat \Bx_{q_1}-\hat \Bx_{q_2}) \cdot \hat \By_0| \le \dfrac{k}{\pi}.\end{equation}
Thus, choosing $\hat \By_0$ outside of the finite number of hyperplanes defined by the $\left(
\begin{array}{c}
 n \\
 2 \\
\end{array}
\right) \left (2 \left \lfloor \dfrac{k}{\pi} \right  \rfloor +1 \right )$ equations of the form \eqref{eqn2}, will give rise to a set of $n$ $\By$-values (i.e., $\By_p = p \hat \By_0$ for $p = \overline{0, n-1}$) forcing the solution $c=(c_1,c_2,...,c_n)$ of \eqref{cj_eqn} to satisfy  $c_j =0$ for all $j=\overline{1,n}$. Therefore, the kernel of $\calD^*$ is trivial and so $\calD$ has a dense range.
\end{proof}

\subsection{Numerical Simulations}
\label{sect:numerics}
In this section we present several numerical simulations supporting the theoretical results presented above. We further develop the scheme proposed in   \cite{Onofrei2014}, \cite{Platt2018} and \cite{Egarguin2018} to accommodate the added constraints on the radiated far field pattern. For a given  $f = (f_1, f_2,..., f_m) \in Y$ and far field pattern values  $c= (c_1, c_2,...,c_n) \in \CC^n$, the problem is to find $w \in L^2(\partial D'_a)$ such that
\begin{equation}
\calD w \approx f \times c.
\label{operatorialeqn}
\end{equation}
To solve \eqref{operatorialeqn}, we employ a method of moments approach by discretizing the control regions into a mesh of collocation points and writing the density $w$ in terms of local basis functions as in \cite{EgarguinWM2020} (see also \cite{Doicu}, \cite{Polycarpou2006}). Hence, the problem is reduced to a linear system of the form 
\begin{equation}
Aw_\text{d}\approx b,
\label{lineareqnMoM}
\end{equation}
where $A$ is the coefficient matrix of dimensions $N_r \times N_c$ where $N_r$ is the total number of mesh points in all near controls and far field directions and $N_c$ is the number of local basis functions used in representing $w$. The vector $w_\text{d}$ of the discrete unknown coefficients of $w$ is computed as the Tikhonov solution
\begin{equation}
\label{c_soln}
w_\text{d} = (\alpha I + A^*A)^{-1}A^*b
\end{equation}
for some optimal regularization parameter $0< \alpha \ll 1$ calculated using the Morozov discrepancy principle, where $A^*$ denotes the complex conjugate transpose of $A$ (see \cite{Platt2018}).

\begin{figure}[!h]
\centering
\includegraphics[width=0.35\textwidth]{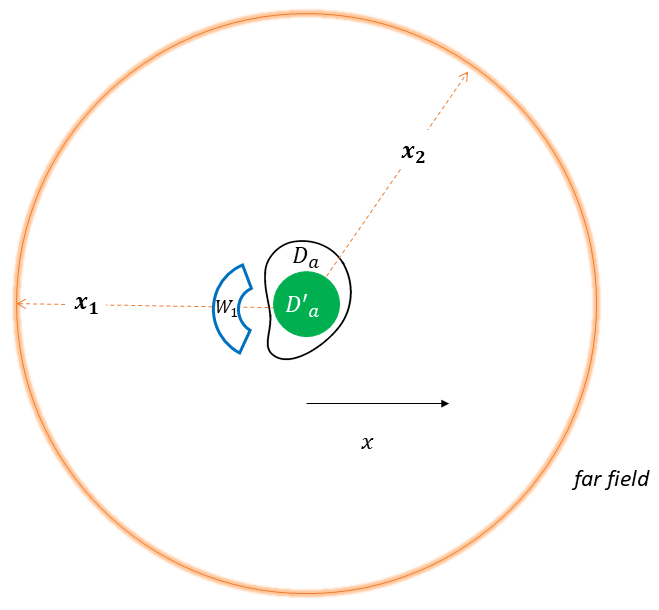}
\caption{Sketch of the top view (plane $z=0$) of the problem geometry showing the near control $W_1$ and the far field directions $\mathbf{x_1}$ and  $\mathbf{x_2}$}
\label{fardir_geom}
\end{figure}

In the following simulations the fictitious source domain $D_a'$ is a sphere of radius $0.01$ m centered at the origin while for simplicity, the physical source domain $D_a$ is chosen to be the sphere of radius $0.015$ m centered at the origin (though in general, it can be any Lipschitz compact domain with $D_a'\Subset D_a$ which does not intersect the near field control regions). We consider the control problem in which the far field direction $\hat{\Bx}_1$ is situated behind a near field. We consider two cases: first, when we prescribe a null in the near field control region $W_1$, hence mimicking communications through an obstacle and second, when the near field is the outgoing plane wave  $f(\mathbf x) = e^{i \mathbf x \cdot (10\mathbf d)}$ with $\mathbf d = \left [-1, 0, 0 \right ]$, simulating covert communication. The unknown density is defined on $\partial D_a'$ by using 234 local basis functions.  The near control is the annular sector \[W_1=\left \{(r, \theta, \phi): r \in [0.02, 0.03],  \theta \in \left [ \frac{\pi}{4}, \frac{3\pi}{4} \right ], \phi \in \left [ \frac{3 \pi}{4}, \frac{5 \pi}{4} \right ] \right \}\] in spherical coordinates with respect to the origin where $r$ is the radius, $\theta \in [0, \pi]$ is the inclination angle and $\phi \in [0, 2 \pi)$ is the azimuthal angle. This sector is discretized into 4640 points. The far field directions in Cartesian coordinates are $\mathbf{\hat x_1} =[-1, 0, 0]$, directly behind the near control and $\mathbf{\hat x_2} =\left [\frac{1}{2}, \frac{1}{2}, -\frac{\sqrt 2}{2} \right ]$. The problem geometry is shown in Figure \ref{fardir_geom}.
To check the accuracy of the generated fields, we provide the plots of the prescribed and generated near fields and when applicable the pointwise relative error. As a further numerical stability check, these fields were plotted in a mesh of points slightly off the set of points used in the collocation scheme. Then aside from stating the generated far field pattern and the relative error, whenever applicable, on the directions $\mathbf {\hat x_1}$ and $\mathbf {\hat x_2}$ we also present the generated far field pattern on a small patch around the two directions. The computed normal velocity on the physical source domain is characterized in two-dimensional $\theta \phi$-plots of its magnitude and real and imaginary parts. We also calculate the average radiated power by the source given by
\begin{equation}
\label{power}
P_{ave} =\dfrac{1}{2}  \int_{\partial B_r(\B0)} \text{ Re} \left [\overline{u} (\nabla u\cdot \Bn) \right ] dS
\end{equation}
where $\partial B_r(\B0)$ is the surface of any sphere containing the source and in our calculations we will evaluate the power in dB relative to a reference level of $10^{-12}$ W.

\subsubsection{A null near field}

\begin{figure}[!h]
\centering
\includegraphics[width=0.5\textwidth]{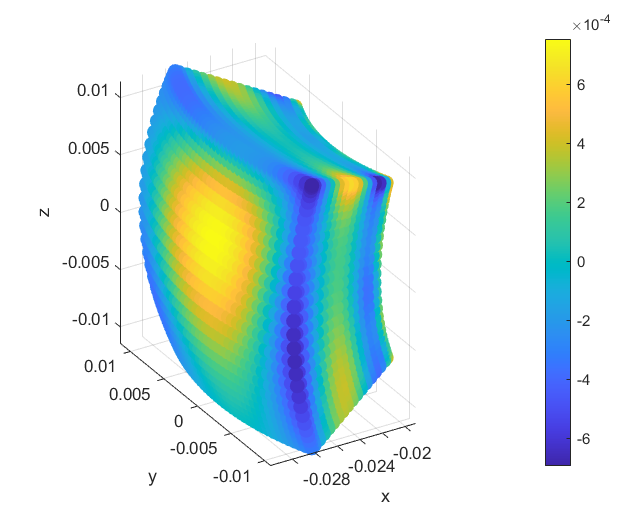}
\caption{Real part of the generated near field}
\label{nearnull2_near}
\end{figure}

In this test, we simulate the case of communicating while avoiding an obstacle and keeping a low signature in another far field direction. We prescribe a null field in $W_1$ and the pair of far field pattern values 0.01 and 0 in the directions of $\mathbf {\hat x_1}$ and $\mathbf {\hat x_2}$, respectively. Figure \ref{nearnull2_near} shows the generated field on the near control. The field on the near control region has maximum pointwise magnitude of about $8 \times 10^{-4}$.

\begin{figure}[!h]
\centering
\subfloat[Generated field on a patch around $\mathbf{\hat x_1}$]{%
\resizebox*{5.5cm}{!}{\includegraphics{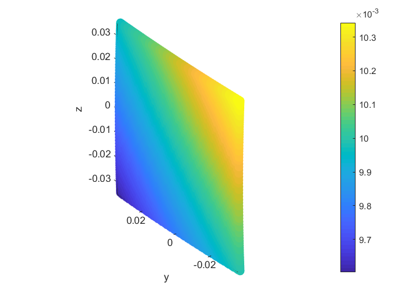}}}\hspace{5pt}
\subfloat[Relative difference from the prescribed value]{%
\resizebox*{5.5cm}{!}{\includegraphics{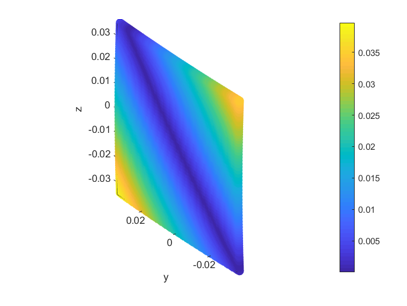}}}\hspace{5pt} \\
\subfloat[Generated field on a patch around $\mathbf{\hat x_2}$]{%
\resizebox*{5.5cm}{!}{\includegraphics{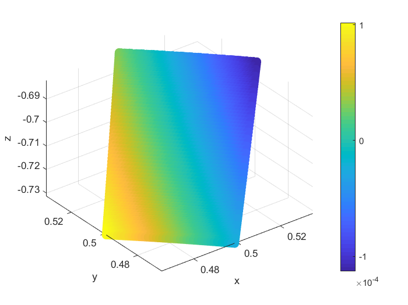}}}\hspace{5pt}
\caption{Results of the far field pattern directional control} 
\label{nearnull2_farfield}
\end{figure}

Figure  \ref{nearnull2_farfield} shows the generated far field pattern values on the two directions. These plots suggest a good approximation of the far field values even on the patches around  $\mathbf {\hat x_1}$ and $\mathbf {\hat x_2}$. The generated pattern value for  $\mathbf {\hat x_1}$ is approximately $0.00998$, with a relative error of only about 0.22\%. Meanwhile the generated value on $\mathbf {\hat x_2}$ is $-1.5711 \times 10^{-6}$.

\begin{figure}[!h]
\centering
\subfloat[Magnitude]{%
\resizebox*{5.29cm}{!}{\includegraphics{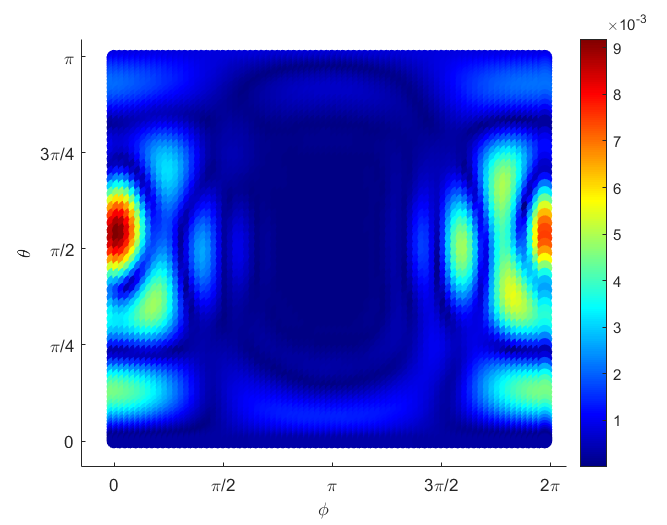}}}\hspace{5pt}
\subfloat[Real part]{%
\resizebox*{5.29cm}{!}{\includegraphics{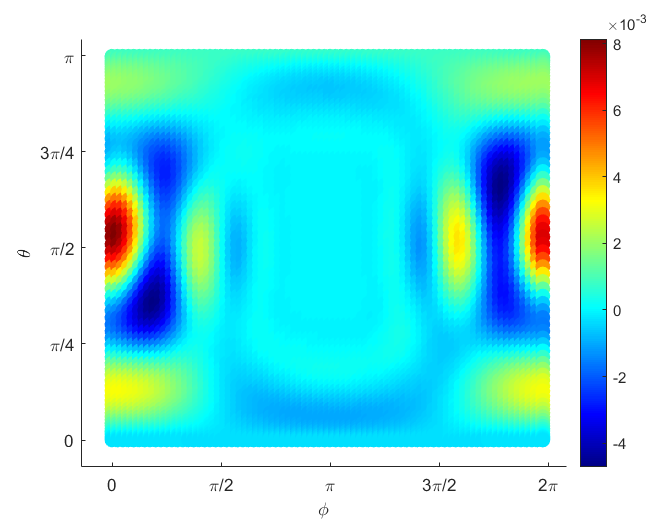}}}\hspace{5pt}
\subfloat[Imaginary]{%
\resizebox*{5.29cm}{!}{\includegraphics{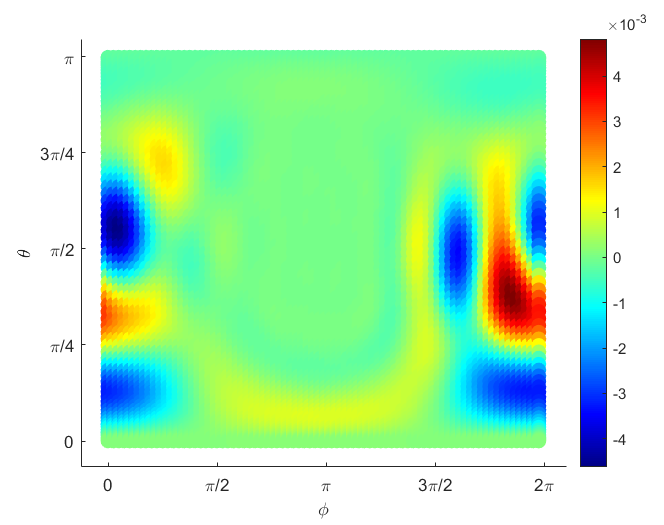}}}\hspace{5pt}
\caption{The computed normal velocity on the actual source} 
\label{nearnull2_nv}
\end{figure}

The computed normal velocity on the physical source is characterized in Figure \ref{nearnull2_nv}. Here, we see that these values has magnitudes values of order $10^{-3}$. The average power radiated by the source is approximately $5.80 \times 10^{-4}$ or around 87.63 dB. 

\subsubsection{A plane wave in the near field}
\begin{figure}[!h]
\centering
\subfloat[Prescribed field]{%
\resizebox*{5.25cm}{!}{\includegraphics{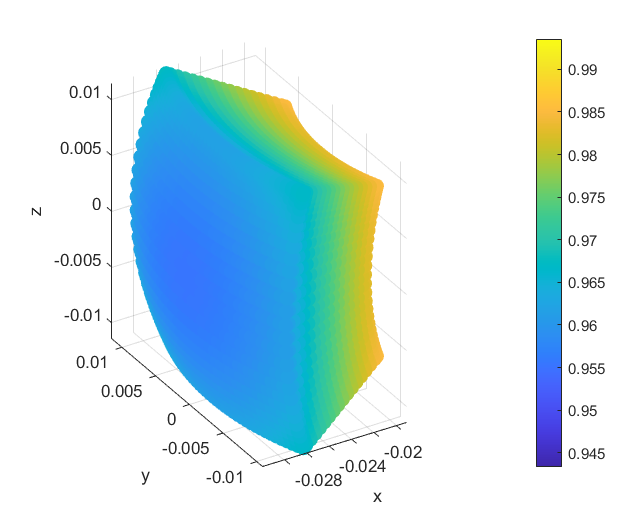}}}\hspace{5pt}
\subfloat[Generated field]{%
\resizebox*{5.25cm}{!}{\includegraphics{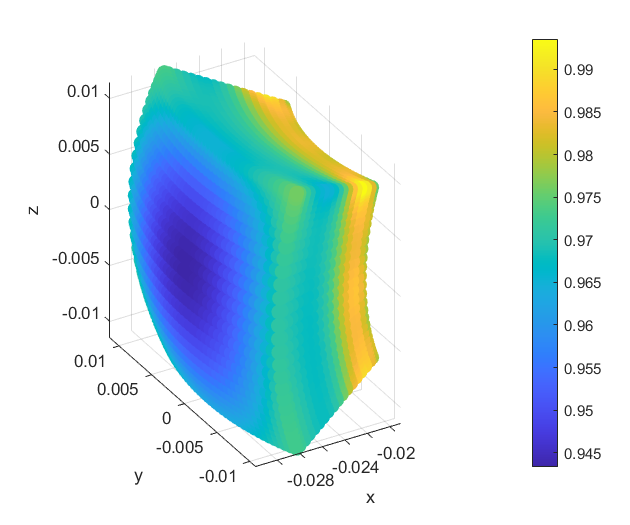}}}\hspace{5pt}
\subfloat[Relative error]{%
\resizebox*{5.25cm}{!}{\includegraphics{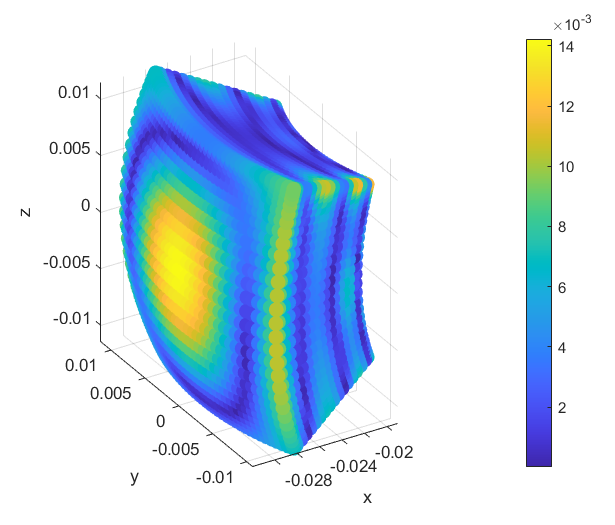}}}\hspace{5pt}
\caption{Results of the field synthesis on the near control} 
\label{nearplane2_near}
\end{figure}

\begin{figure}[!h]
\centering
\subfloat[Generated field on a patch around $\mathbf{\hat x_1}$]{%
\resizebox*{5.5cm}{!}{\includegraphics{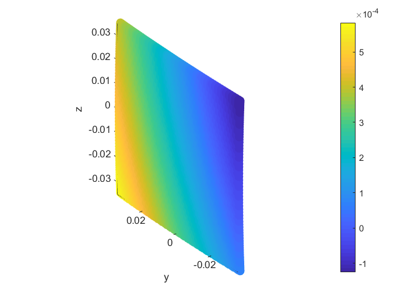}}}\hspace{5pt}\\ 
\subfloat[Generated field on a patch around $\mathbf{\hat x_2}$]{%
\resizebox*{5.5cm}{!}{\includegraphics{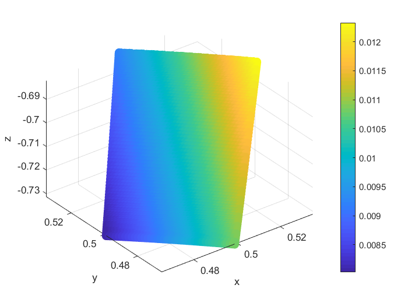}}}\hspace{5pt}
\subfloat[Relative difference from the prescribed value]{%
\resizebox*{5.5cm}{!}{\includegraphics{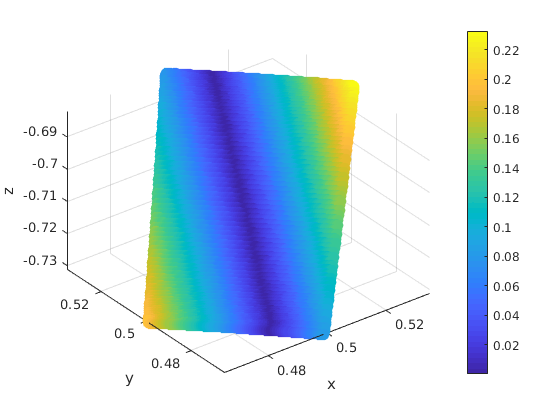}}}\hspace{5pt}
\caption{Results of the far field pattern directional control} 
\label{nearplane2_farfield}
\end{figure}

In this experiment, we synthesize a plane wave on the near control while keeping the direction behind it a quiet zone and projecting a pattern in another far field direction. We prescribe the left traveling plane wave $f(\mathbf x) = e^{i \mathbf x \cdot (10\mathbf d)}$ with $\mathbf d = \left [-1, 0, 0 \right ]$ on the near control, a zero far field pattern value in the direction $\mathbf {\hat x_1}$ behind it and 0.01 in the direction $\mathbf {\hat x_2}$.  The results of the near field approximation is shown in Figure \ref{nearplane2_near}. The near field relative errors do not exceed 1.5 \%.
The results of the far field pattern synthesis are shown in Figure \ref{nearplane2_farfield}. The results on the patch around $\mathbf {\hat x_1}$ are good, with generated values  of order $10^{-4}$. In particular, the generated value at $\mathbf{\hat x_1}$ is around $2.6 \times 10^{-4}$. Meanwhile on the patch around  $\mathbf{\hat x_2}$, there are points where the relative error reach 22\%. But this decreases to desirable values for points near  $\mathbf{\hat x_2}$. In fact, the generated value at $\mathbf{\hat x_2}$ is $0.01004$ with a relative error of just about 0.021\%

\begin{figure}[!h]
\centering
\subfloat[Magnitude]{%
\resizebox*{5.29cm}{!}{\includegraphics{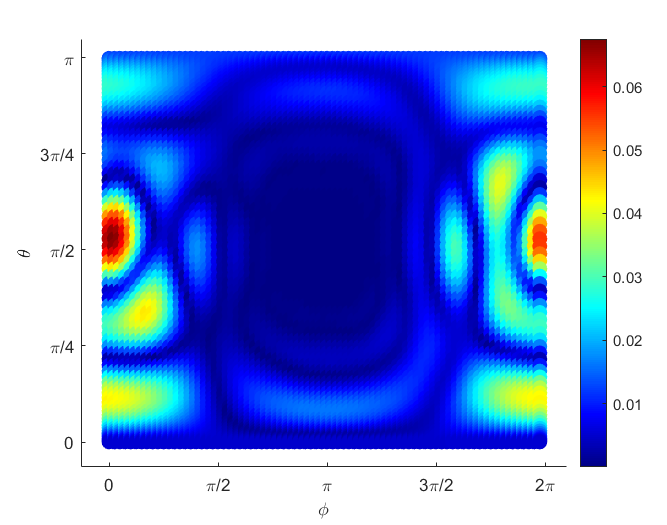}}}\hspace{5pt}
\subfloat[Real part]{%
\resizebox*{5.29cm}{!}{\includegraphics{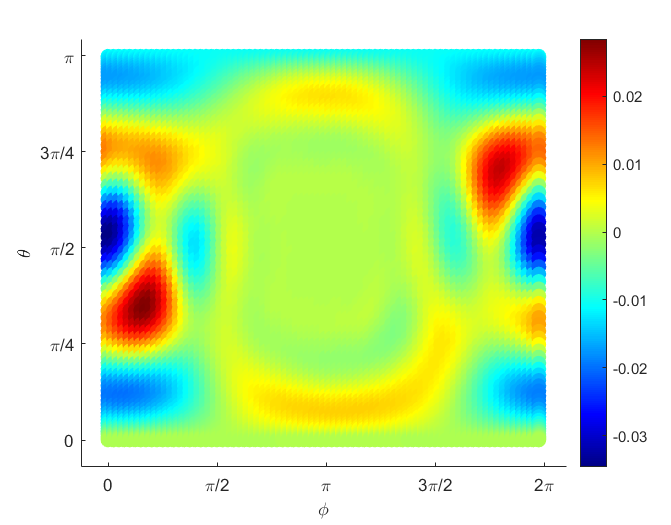}}}\hspace{5pt}
\subfloat[Imaginary]{%
\resizebox*{5.29cm}{!}{\includegraphics{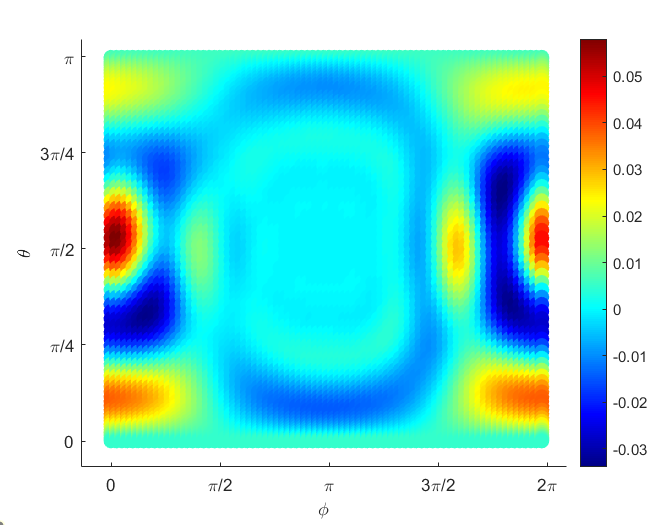}}}\hspace{5pt}
\caption{The computed normal velocity on the actual source} 
\label{nearplane2_nv}
\end{figure}

Lastly, we look at the calculated normal velocity. Figure \ref{nearplane2_nv} displays the pointwise magnitude as well as the real and imaginary parts of the normal velocity on the physical source. The average acoustic power radiated by the source is around $2.44 \times 10^{-2}$ or about 103.87 dB which, as expected is larger than in the previous simulations due to the extra work the source needs to  do now to create a plane wave in the near field control region.

\section{Homogeneous Ocean Environment}
\label{oceanenv}

In this section, we prove the possibility of near field active control while maintaining desired radiation in prescribed far field directions in a homogeneous ocean environment of constant depth. The problem is similar to the one presented in Section \ref{freespace} except that the sources and the control regions are submerged in a homogeneous ocean environment. The near field control problem was briefly discussed in \cite{Platt2018} without numerical simulations. Aside from providing numerical validation, this section adds the novelty of incorporating additional far field pattern constraints in the theoretical analysis which is an important feature from the point of view of applications and theoretically nontrivial in this particular environment.

\begin{figure}[!h]
\centering
\includegraphics[width=0.75\textwidth]{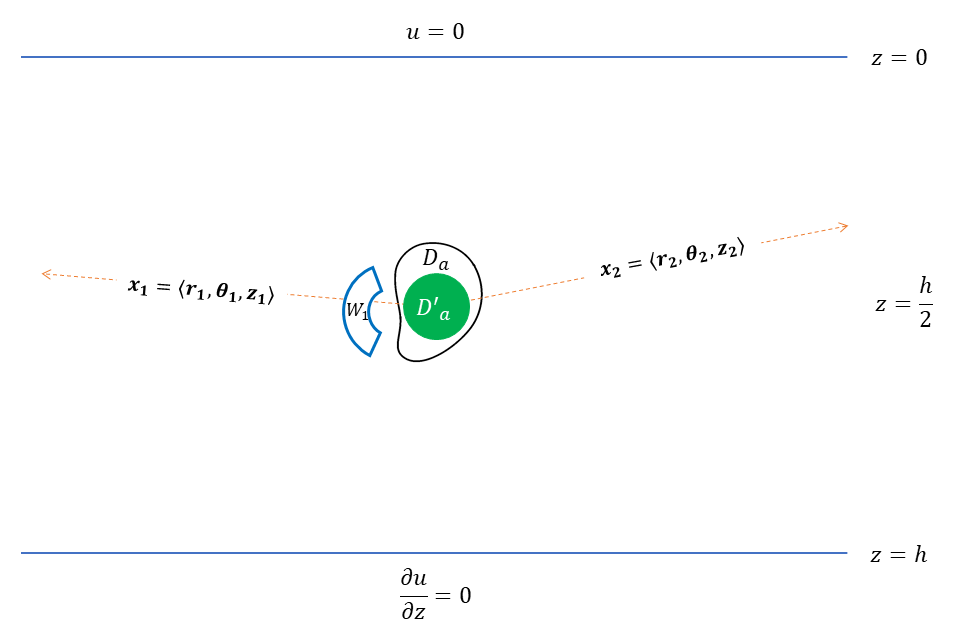}
\caption{A 3D sketch of the problem geometry for the homogeneous ocean showing the near control and two far field directions $\mathbf{x_1}$ and  $\mathbf{x_2}$}
\label{ocean}
\end{figure}

Assuming the same notations as in the theoretical set-up of Section \ref{freespace} the problem is modeled by \eqref{P1a}, \eqref{P1b} with the boundary conditions \eqref{bc-ocean0}
\begin{equation}
\begin{cases}
u &=0 \text{ at the ocean surface } z=0 \text{ and } \\
 \dfrac{\partial u}{\partial z} &= 0 \text{ at the ocean floor }z=h.
 \end{cases}
\end{equation}
and the radiation condition described below at \eqref{HP_ocean}. A sketch of the geometry is shown in Figure \ref{ocean}.
We continue with the presentation of the theoretical framework and results. Then we perform some numerical simulations that illustrate the feasibility of the proposed theoretical and numerical framework.
 
\subsection{Theoretical Framework}
\label{ocean_theory}
The mathematical and numerical framework from our previous works can be adapted for the homogeneous ocean environment. In this section we will assume that the entire functional framework (notations, geometrical conditions and functional assumptions) formulated in Section \ref{problem} remains the same for the case of homogeneous oceans of constant depth unless otherwise specified. The major adjustment is the Green's function for this new medium. The corresponding Green's function for this environment has the normal mode representation for an evaluation point $\mathbf x = (r, \theta, z)$ and source point $\mathbf y = (0, \theta_0, z_0)$  in cylindrical coordinates
\begin{equation}
\label{Green-ocean}
G(\mathbf x, \mathbf y) = \dfrac{i}{2h} \sum_{p=0}^{+\infty} \phi_p(z) \phi_p(z_0) H_0^{(1)}(ka_pr)
\end{equation}
where $H_0^{(1)}$ is the Hankel function of order zero of the first kind, $\phi_p$ is the $p^{\text{th}}$ modal solution with associated eigenvalue $a_p$ (see \cite{Keller1997}, \cite{MarineAcoustics}, \cite{Kuperman2011}). These eigenpairs are given by
\begin{equation}
\label{a_n}
a_p = \sqrt{1-\dfrac{(2p+1)^2\pi^2}{4k^2h^2}} \text{ and}
\end{equation}
\begin{equation}
\label{phi_n}
\phi_p(z) = \sin \left ( k \sqrt{1-a_p^2}z \right) = \sin \left (\frac{(2p+1) \pi}{2h} z \right).
\end{equation}
As proved in \cite{MarineAcoustics}, the function $G$ can be expressed as a continuous perturbation of the Green's function in free space. We will assume that the physical source $D_a$ satisfy $\Bx\cdot \Bn\geq 0$ for any $\Bx\in \partial D_a$ where $\Bn$ denotes the exterior normal to $\Bx\in \partial D_a$. 

With these notations, the forward problem in the homogeneous finite-depth ocean environment $R= \{\Bx = (r, \theta ,z) \in \RR^3~|~ z \in [h, 0]\}$ can be formulated as follows: for a given boundary input $u_b$ on the surface of the source $\partial D_a$ find  $u$ solution of the following exterior Helmholtz problem 
\begin{equation}
\vspace{0.15cm}\left\{\vspace{0.15cm}\begin{array}{llll}
\GD u+k^2u=0 \mbox{ in } R \!\setminus D_a \vspace{0.15cm},\\
u=u_b \mbox{ on }\partial D_a\\
u=0 \text{ at the ocean surface } z=0 \text{ and } \\
\dfrac{\partial u}{\partial z} = 0 \text{ at the ocean floor }z=h \\
\displaystyle \lim_{r \to \infty} r^{1/2} \left (\dfrac{\partial u_p}{\partial r} -ika_pu_p\right ) = 0, \text{uniformly for all } \theta \in [0, 2 \pi) , 
\end{array}\right.
\label{HP_ocean}
\end{equation}
where $u_p$ in the radiation condition above are normal modes appearing in the representation of the solution $u$, i.e., 
\begin{equation}
\label{normal-mode-uo} 
u(\Bx)=\sum_{p=0}^{\infty} \phi_p(z)u_p(r,\theta), \mbox{ for } r \mbox{ large enough}.
\end{equation}

Classical manipulations and the definition of the Green's function introduced at \eqref{Green-ocean} imply that, for any density $w \in C(\partial D'_a)$, the following function
\begin{equation}
\label{sl-ocean}
u(\Bx) = \int_{\partial D'_a}  w(\By) G(\Bx, \By) dS_\By
\end{equation}
is a solution to \eqref{HP_ocean} with $u_b\in C^{\infty}(\partial D_a)$ given by $u_b=\int_{\partial D'_a}  w(\By) G(\Bx, \By) dS_\By$.  Note that, as before, we make use of a fictitious spherical source domain $D_a'$ to ensure smoothness of our boundary input $u_b$. It was shown in \cite{MarineAcoustics} that for a given density $w \in \partial D_a'$, $u$ defined above has an asymptotic representation given by
\begin{equation}
u(\Bx) = \sum_{p=0}^N \dfrac{1}{\sqrt{ka_pr}} e^{ika_pr}g_p(\theta, z) + \mathcal O(\dfrac{1}{r^{3/2}}), \text{ as } r \to +\infty
\end{equation}
where $\Bx =(r, \theta, z)$ (in cylindrical coordinates) and for each $p= \overline{0, N}$ and $\By = (r', \theta', z') \in \partial D_a'$ in cylindrical coordinates,
\begin{equation}
g_p( \theta, z) = \sqrt{\dfrac{2}{\pi}} \int_{\partial D_a'}  w(\By) \left ( \sum_{q=0}^{\infty} e^{-i(q+\frac{1}{2})\frac{\pi}{2}} \alpha_{qp}(z, \theta, r', z', \theta')\right ) dS_\By
\end{equation}
and 
\begin{equation}
\alpha_{qp}(z, \theta, r', z', \theta') = \dfrac{i \epsilon_q}{2h} \phi_p(z) \left [ \cos(q \theta) \beta_{qp}(\By) + \sin(q \theta) \gamma_{qp}(\By) \right]
\end{equation}
where $\epsilon_0=1$, and $\epsilon_q=2$ for $q\geq 1$ and where 
\begin{align}
\beta_{qp}(\By) &=J_q(ka_pr')\phi_p(z')\cos q \theta' \text{ and} \\
\gamma_{qp}(\By) &= J_q(ka_pr')\phi_p(z')\sin q \theta'.
\end{align}
In the last two equations above $J_q$ is the Bessel function of the first kind of order $q$. Then following \cite{MarineAcoustics} we define the far field pattern as the function $u_\infty$ given by
\begin{equation}
\label{sl-ff}
u_\infty(\hat \Bx) = \sum_{p=0}^N g_p( \theta, z),
\end{equation}
where $\hat \Bx =(1, \theta, z)$ and $N > \frac{kh}{\pi}-\frac{1}{2}$  so that the terms $g_p$ removed from the sum are all evanescent (non-propagating) modes.  

\begin{rem}
	\label{int-ctr-ocean}
		\label{int-ctr-ocean}
	The restriction that each $f_l$ satisfies the Helmholtz equation in some neighborhood of $R_l$ and the fact that $R_l \Subset W_l$ for all $1\leq l\leq m$ ensure, through uniqueness and regularity results for the interior Helmholtz problems (in the spirit of \cite{Onofrei2014}), that the field $u$, solution of \eqref{HP_ocean}, will satisfy the control constraint \eqref{P1b} if 
	$$
	\begin{cases}
	\Vert u-f_l\Vert_{L^2(\partial W_l)}\leq \mu \text{ for } l= \overline{1,m} \\
	|u_\infty(\hat \Bx_j)-f_{\infty,j}| \leq \mu \text{ for } j= \overline{1,n}
	\end{cases}.
	$$
\end{rem}

Hence, from the Remark \ref{int-ctr-ocean} we deduce that the control problem \eqref{HP_ocean}, \eqref{P1b} amounts to to finding the density $w \in L^2(\partial D'_a)$ so that the corresponding solution $u$ of  \eqref{HP_ocean} and its corresponding far field pattern $u_\infty$ satisfy
\begin{equation}
\begin{cases}
\Vert u - f \Vert_{L^2(\bigcup_{l =1}^m \partial W_l)}\leq \mu \\
|u_\infty(\hat \Bx_j)-f_{\infty,j}| \leq \mu \text{ for } j= \overline{1,n}
\label{farfieldocean}
\end{cases}
\end{equation}
for any  $f = (f_1, f_2,..., f_m) \in Y$ and fixed directions $\hat \Bx_j$, $j =\overline{1,n}$.
 Such a density will give us then the necessary source presure characterization $u_b=\int_{\partial D'_a}  w(\By) G(\Bx, \By) dS_\By$ on the physical source $D_a$ so that its radiated field satisfies the required control conditions \eqref{P1b}.  

Because the control problem is again reduced to finding the density on the surface of the fictitious source $D_a'$ we note that the same machinery developed in the previous section can be employed after the making the appropriate modification of the Green's function.

In parallel to the notations in the previous section, we define the near field propagator operator $\calK: L^2(\partial D'_a) \to Y$, by
\begin{equation}
\calK w (\By_1, \By_2,..., \By_m) = \big (\calK_1 w (\By_1), \calK_2 w (\By_2), ..., \calK_m w (\By_m) \big )
\end{equation}
where for each $l =\overline{1,m}$, $\By_l \in W_l$
\begin{equation}
\calK_l w (\By_l) = \displaystyle \int_{\partial D'_{a}}w(\Bx) G(\Bx,\By_l) dS_\Bx.
\end{equation}
and for each far field direction with cylindrical coordinates $\hat \Bx_j= (1, \theta_j, z_j)$, $j = \overline{1, n}$ with $z_j\in(h,0)$, $\theta_j \in [0, 2 \pi)$ and $\theta_{j_1} \ne \theta_{j_2}$ for $j_1 \ne j_2$, we define the far field pattern propagator $\calK_{\infty}: L^2(\partial D'_a) \to \CC^n$ as \[\calK_\infty w = \big ( \calP_{w,1}, \calP_{w,2}, ..., \calP_{w,n} \big )\] where
\begin{equation}
\calP_{w,j}  =  \sum_{p=0}^N g_p( \theta_j, z_j).
\end{equation}
Finally, we define the operator $\calD: L^2(\partial D'_a) \to Y \times \CC^n$ such that
\begin{equation}
\calD w (\By_1, ..., \By_m) = \big ( \calK_1 w (\By_1), ..., \calK_m w (\By_m),   \calP_{w,1}, \calP_{w,2}, ..., \calP_{w,n} \big ).
\label{propagator_ocean}
\end{equation}
To prove that the range of the linear compact operator $\calD$ is dense in $Y\times \CC^n$, i.e., any target in $Y\times \CC^n$ can be approximated by an image under $\calD$, we shall show in the next theorem that the adjoint operator $\calD^*$ has a trivial kernel.

\begin{thm}
Except a discrete set of values for $k$, the operator $\calD$ defined in \eqref{propagator_ocean} has a dense range.
\end{thm}

\begin{proof}
Again, we prove the equivalent statement that $\calD^*$ has a trivial kernel. To do so, we adapt the arguments used in the proof of Theorem \ref{thm:free}. Straightforward calculations will show that the adjoint operator $\calD^*:Y \times \CC^n \to  L^2(\partial D'_a)$ is given by
\begin{equation}
(\calD^*(\psi, c))(\By) =  \sum_{l =1}^m \calK^*_l \psi_l (\By) + \sum_{j=1}^n c_j h_j(\By),
\end{equation}
where $\calK^*_l: L^2(\partial W_l) \to L^2(\partial D'_a)$ is given by \[ \calK^*_l \psi_l (\By)  = \int_{\partial W_l} \psi_l(\Bx)\overline G(\Bx, \By) dS_\Bx. \] for any $\By \in \partial D'_a$ and $h_j: \partial D'_a \to \CC$ is defined as \[ h_j(\By) = \sqrt{\dfrac{2}{\pi}} \sum_{p=0}^N \sum_{q=0}^\infty e^{i(q+\frac{1}{2})\frac{\pi}{2}} \overline \alpha_{qp} (z_j,  \theta_j, \By).\]
Consider $(\psi, c)\in Y \times \CC^n $ with $(\calD^*(\psi, c))=0$. Let
\begin{equation}
\label{oceans-int}
w(\By)= \int_{\partial W_l} \overline \psi_l(\Bx)G(\Bx, \By) dS_\Bx+ \sum_{j=1}^n \overline c_j \overline h_j(\By),
\end{equation}
It is simple to observe that $w$ defined in\eqref{oceans-int} is a solution of the interior Helmholtz equation in $D_a'$ with zero Dirichlet data on the boundary (since by definition $\calD^*(\psi, c)=0$), and except a finite set of values for $k$ this implies that $w=0$ in $D_a'$. Next, in the same spirit as we did for the case of free space environments in the previous section, using analytic continuation and the same continuity and jump relations for the single layer potential used in the proof of Theorem \ref{thm:free} (which still apply since $G$ is a continuous perturbation of the Green's function in free space), we discern that $w(\By)=0$,  for $\By\in R$. This, and the jump conditions for the single layer potential (which still apply since $G$ is a continuous perturbation of the Green's function in free space) imply $\psi_l = 0$ on $\partial W_l, l=\overline{1, m}$. Thus, using this in $w(\By)=0$,  for $\By\in R$ recalling \eqref{oceans-int} we obtain the following condition for $c=(c_1,c_2,...c_n)$:
\begin{equation}
 \sum_{j=1}^n  \sum_{p=0}^N \sum_{q=0}^\infty \overline c_j e^{-i(q+\frac{1}{2})\frac{\pi}{2}}  \alpha_{qp} ( z_j, \theta_j,\By) =0 \text{ for any $\By \in R$.}
 \label{eqnocean1}
\end{equation}
To prove that the kernel of $\calD^*$ is trivial it remains to show that \eqref{eqnocean1} implies that all $c_j$'s are zero. Let $q_0$ with $0 \le q_0 < \infty$ be arbitrarily fixed. Taking the inner product of both sides of \eqref{eqnocean1} against $\cos q_0 \theta'$, applying the orthogonality property of sines and cosines and algebraic manipulations yields for any $\By = (r', \theta', z') \in R$
\begin{equation}
 \sum_{j=1}^n  \sum_{p=0}^N  \overline c_j \phi_p(z_j) J_{q_0}(ka_pr') \phi_p(z') \cos (q_0 \theta_j)=0.
 \label{eqnocean2}
\end{equation}
Note that \[\dfrac{d^{(l)}}{dz'} \phi_p(0) = \begin{cases} 0, & \text{ if $l$ is even} \\ k^l (1-a_p^2)^{l/2}, & \text{ for } l=1, 5, 9,... \\ -k^l (1-a_p^2)^{l/2}, & \text{ for } l=3, 7, 11,...\end{cases}.\] Define $A_p = \ds \sum_{j=1}^n \overline c_j \phi_p(z_j) \cos(q_0 \theta_j)$ and let $B_p= k(1-a^2_p)^{1/2}$. Taking the the order $l=1+4s$ derivative of both sides of \eqref{eqnocean2} with respect to $z'$ evaluated at $z'=0$, one obtains the system
\begin{equation}
\sum_{p=0}^N  A_p  J_{q_0}(ka_pr')  B_p^{1+4s}=0, s = \overline{0, N}.
 \label{eqnocean3}
\end{equation}
Letting $\lambda_p = B_p^4$, system \eqref{eqnocean3} can be viewed as an $N+1 \times N+1$ system with unknowns $A_pB_p$ with coefficient matrix
\begin{equation}
 D=\left[
\begin{array}{cccc}
 J_{q_0}(ka_0r') & J_{q_0}(ka_1r') & \cdots  & J_{q_0}(ka_Nr') \\
 J_{q_0}(ka_0r')  \lambda_0 & J_{q_0}(ka_1r') \lambda_1 & \cdots  & J_{q_0}(ka_Nr') \lambda_N \\
   &   & \vdots  &   \\
 J_{q_0}(ka_0r')  \lambda^N_0 & J_{q_0}(ka_1r') \lambda^N_1 & \cdots  & J_{q_0}(ka_Nr') \lambda^N_N \\
\end{array}
\right] 
\label{eqnocean4}
 \end{equation}
with $\det D = \left (\ds \prod_{p=0}^N J_{q_0}(ka_pr') \right ) \left ( \ds \prod_{1 \le p < l \le n} (\lambda_p-\lambda_l) \right )$. Note that by definition $\lambda_p - \lambda_l \ne 0$ for $p \ne l$. From \cite{Breen1995}, the smallest root of $J_{q_0}$ is  bounded below by $q_0 + \frac{2}{3} |\xi|^{3/2}$, where $\xi = -0.36605...$ is the smallest negative root of the Airy function. Since the $a_p$'s are decreasing then choosing $r'$ so that
\[r' < \dfrac{q_0 + \frac{2}{3} |\xi|^{3/2}}{ka_0}  \] 
makes $\det D \ne 0$. Hence, \eqref{eqnocean3} only has the trivial solution $A_pB_p=0$  for all $p  = \overline{0, N}$ implying
\begin{equation}
\ds \sum_{j=1}^n \overline c_j \phi_p(z_j) \cos(q_0 \theta_j) =0.
\end{equation}
On the other hand, taking the inner product of both sides of \eqref{eqnocean1} against $ \sin q_0 \theta'$ and doing analogous calculations as above yields
\begin{equation}
\ds \sum_{j=1}^n \overline c_j \phi_p(z_j) \sin(q_0 \theta_j) =0
\end{equation}
 for all $p  = \overline{0, N}$. In particular for $p=0$, the last two equations imply
 \begin{equation}
\ds \sum_{j=1}^n \overline c_j \phi_0(z_j) e^{iq_0 \theta_j} =0.
\label{lastsyst}
\end{equation}
Since $q_0$ was arbitrarily chosen, by using the values $q_0=0, 1,..., n-1$ above we obtain the following homogeneous linear system of equations in unknowns $\overline c_j$ with coefficient matrix
\begin{equation}
 E=\left[
\begin{array}{cccc}
 \phi_0(z_1) & \phi_0(z_2)  & \cdots  &  \phi_0(z_n)  \\
 \phi_0(z_1)e^{i \theta_1} & \phi_0(z_2) e^{i \theta_2} & \cdots  &  \phi_0(z_n)e^{i \theta_n} \\
   &   & \vdots  &   \\
 \phi_0(z_1)e^{i (n-1)\theta_1} & \phi_0(z_2) e^{i (n-1) \theta_2} & \cdots  &  \phi_0(z_n)e^{i (n-1) \theta_n} \\
\end{array}
\right].
\label{eqnocean5}
 \end{equation}
 Note that $E$ is another Vandermonde-type matrix with determinant \[\det E =  \left (\ds \prod_{j=1}^n \phi_{0}(z_j) \right ) \left ( \ds \prod_{1 \le q < l \le n} (e^{i\theta_q}-e^{i \theta_l}) \right ).\]
 This will be zero if and only if there exists a $z_j$ such that $\phi_0(z_j)=0$ or equivalently, $z_j = 2t h$ for some integer $t$. However, this cannot be the case since $z_j \in (h, 0)$. Hence, \eqref{lastsyst} has a unique solution, namely $c_j =0, j=\overline{1,n}$. Therefore, $\ker \calD^*$ is trivial and consequently, $\calD$ has a dense range.
\end{proof}

\subsection{Numerical Simulations}
\begin{figure}[!h]
\centering
\includegraphics[width=0.5\textwidth]{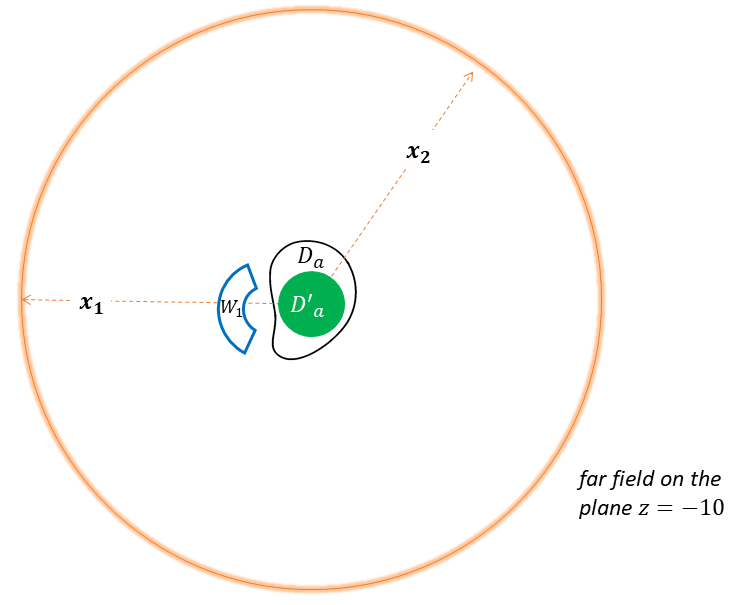}
\caption{A cross section of the simulation geometry}
\label{oceansimulation}
\end{figure}

In this section we present numerical simulations illustrating the results obtained in Section \ref{ocean_theory}. The numerical framework is an adaptation of the one discussed in Section \ref{sect:numerics} where the calculation of the matrix of moments is modified with the corresponding Green's function and far field pattern for the homogeneous oceans environment. To our knowledge, this paper is the first instantiation of numerical simulation support for control problems of the form \eqref{HP_ocean}, \eqref{farfieldocean}. We again consider a near control region $W_1$ and far field directions $\hat \Bx_1 =(1, \theta_1, \frac{h}{2} )$ and $\hat \Bx_2 =\left (1, \theta_2, \frac{h}{2} \right )$.  The control problem is to find the density on the fictitious source $w \in C(\partial D_a')$ such that for a prescribed field $f_1 \in L^2(\partial W_1)$ and prescribed far field patterns $f_\infty(\hat \Bx_j) \in \mathbb C, j=1,2$ the following hold:
\begin{equation}
\begin{cases}
u & \approx f_1 \text{ in } W_1 \\
u_\infty(\hat \Bx_j) & \approx  f_\infty(\hat \Bx_j)
\end{cases}.
\end{equation}
where $u$ and $u_\infty$ are defined at \eqref{sl-ocean} and respectively \eqref{sl-ff}.
In the last simulation, we will add another control $W_2$ where we will prescribe a null field. In all simulations, we consider $h=-20$ m , $k=10$, $n=100$ and $m=100$. The unknown density $w$ is expressed in terms of 234 local basis functions. The fictitious source is a sphere of radius $0.01$ m centered at $(0,0, -10)$ while the actual source is the sphere of radius $0.015$ m with the same center. The near control is the annular sector \[W_1=\left \{(r, \theta, \phi): r \in [0.02, 0.03],  \theta \in \left [ \frac{\pi}{4}, \frac{3\pi}{4} \right ], \phi \in \left [ \frac{3 \pi}{4}, \frac{5 \pi}{4} \right ] \right \},\]  and for the last simulation, we have the null control region \[W_2 =\left \{(r, \theta, \phi) : r \in [0.15, 0.2],  \theta \in \left [\frac{\pi}{4}, \frac{3\pi}{4} \right ] , \phi \in \left [-\frac{\pi}{4}, \frac{\pi}{4} \right] \right \},\] both discretized into 4640 collocation points. For simplicity of notations, $W_1$ and $W_2$ were given in spherical coordinates $(r, \theta, \phi)$, where $r$ is the radius, $\theta \in [0, \pi]$ is the inclination angle and $\phi \in [0, 2 \pi)$ is the azimuthal angle. On the other hand, for consistency with the theoretical framework from the previous section, the far field directions will be given in cylindrical coordinates $(r, \theta, z)$. In the simulations to follow, the far field directions are $\mathbf{\hat x_1} =\left(1, \pi, -10 \right)$ directly behind the near control and $\mathbf{\hat x_2} =\left (1, \frac{\pi}{4}, -10 \right )$. A cross section along the middle plane $z=\frac{h}{2}$ of the simulation geometry is shown in Figure \ref{oceansimulation}.

As before, we present plots of the prescribed and generated fields on the control region/s for a visual comparison of field pattern. The fields were plotted in a mesh of points slightly off the original mesh used for the collocation scheme as a numerical stability test. Whenever applicable, we also plot the pointwise relative errors. The computed normal velocity on the actual source will be characterized by 2D plots of its magnitude, real and imaginary parts in a $\theta \phi$-mesh. We will further describe this surface input by calculating the actual source's average radiated power as defined in \eqref{power}.

\subsubsection{A null near field}
\begin{figure}[!h]
\centering
\subfloat[Using the default color scheme]{%
\resizebox*{7.5cm}{!}{\includegraphics{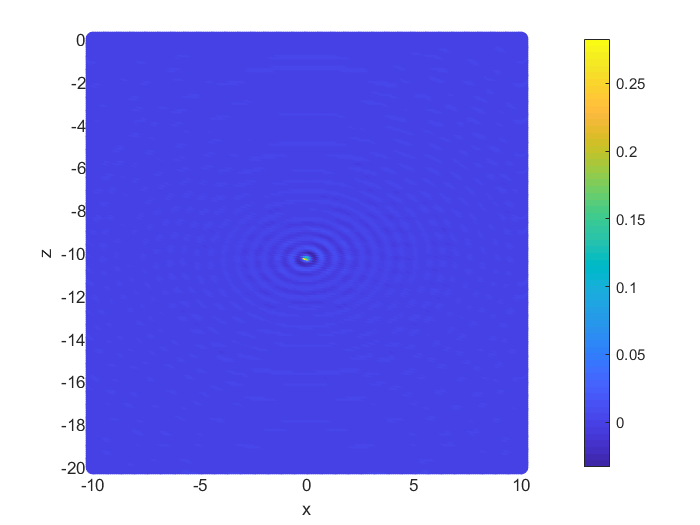}}}\hspace{5pt}
\subfloat[Using a truncated color scheme highlighting the reflections from the boundaries ]{%
\resizebox*{7.5cm}{!}{\includegraphics{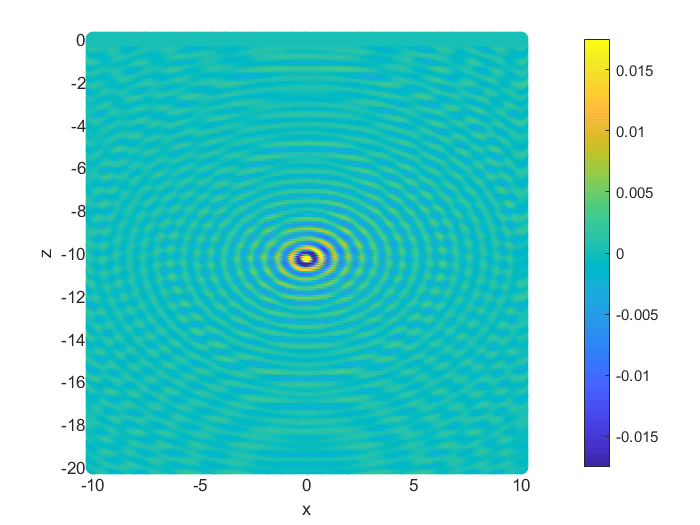}}}\hspace{5pt} \\
\caption{Real part of the generated field on the vertical cross section $y=0$} 
\label{fullplane}
\end{figure}

In this test, we prescribe a null field on $W_1$ and the far field pattern values 0.01 at $\hat \Bx_1$ and 0 at $\hat \Bx_2$. This is a simulation of obstacle-avoiding communication while projecting a quiet zone in a far field direction. The real part of the generated field on the vertical cross section $y=0$ is shown in Figure \ref{fullplane}. The left plot shows the field using the default color bar capturing the entire range of field values. The radiating character of the field is noticeable albeit the very low field values. The plot on the right uses a truncated color bar to reveal the reflections due to the top and bottom ocean boundaries. 

\begin{figure}[!h]
\centering
\includegraphics[width=0.5\textwidth]{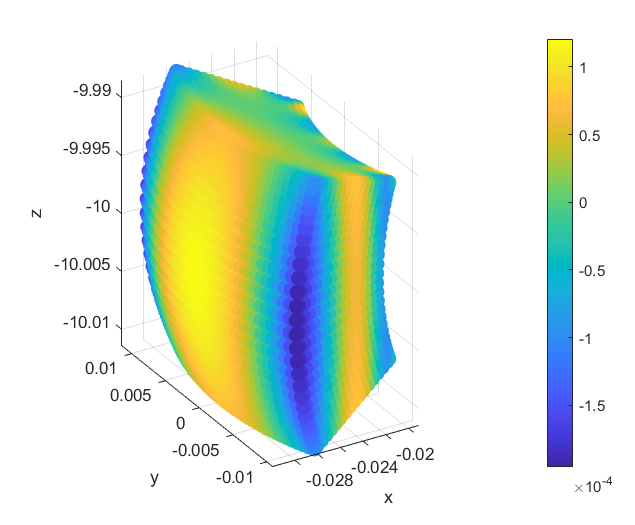}
\caption{Real part of the generated near field}
\label{ocean2_near_re_u}
\end{figure}

The generated near field in the control region $W_1$ is shown in Figure \ref{ocean2_near_re_u}. It can be observed that indeed a low signature was generated in $W_1$ as the field values' magnitude do not exceed $1.96 \times 10^{-4}$.

\begin{figure}[!h]
\centering
\subfloat[Generated field on a patch around $\mathbf{\hat x_1}$]{%
\resizebox*{5.5cm}{!}{\includegraphics{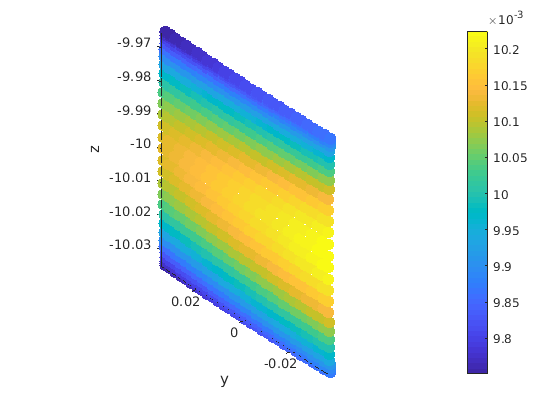}}}\hspace{5pt}
\subfloat[Relative difference from the prescribed value]{%
\resizebox*{5.5cm}{!}{\includegraphics{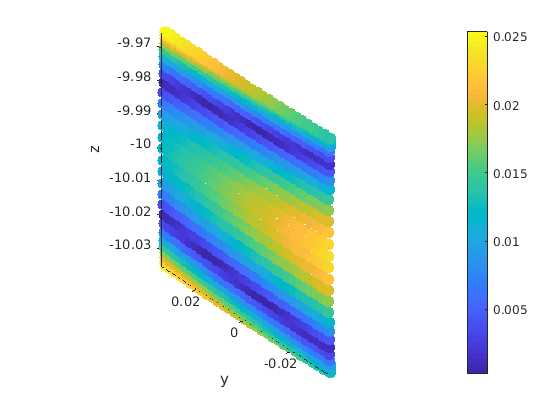}}}\hspace{5pt} \\
\subfloat[Generated field on a patch around $\mathbf{\hat x_2}$]{%
\resizebox*{5.5cm}{!}{\includegraphics{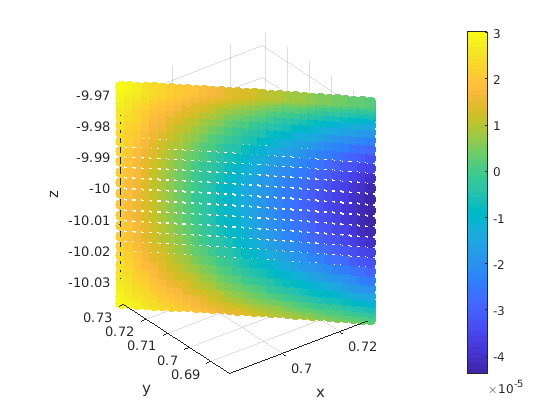}}}\hspace{5pt}
\caption{Results of the far field pattern directional control} 
\label{ocean2_farfielddir}
\end{figure}

The generated far field pattern values on some patches around the two fixed directions are shown in Figure \ref{ocean2_farfielddir}. Around $\hat \Bx_1$, the relative errors reach as high as $2.54\%$. At $\hat \Bx_1$ the generated value is about $0.0102$ with relative error of just $1.78\%$. Around $\hat \Bx_2$, the values has order $10^{-4}$. At $\hat \Bx_2$, the generated value is about $-1.44 \times 10^{-5}$.

\begin{figure}[!h]
\centering
\subfloat[Magnitude]{%
\resizebox*{5.29cm}{!}{\includegraphics{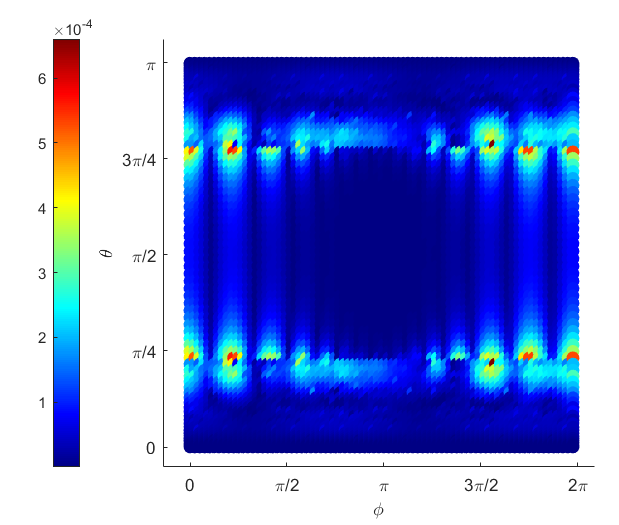}}}\hspace{5pt}
\subfloat[Real part]{%
\resizebox*{5.29cm}{!}{\includegraphics{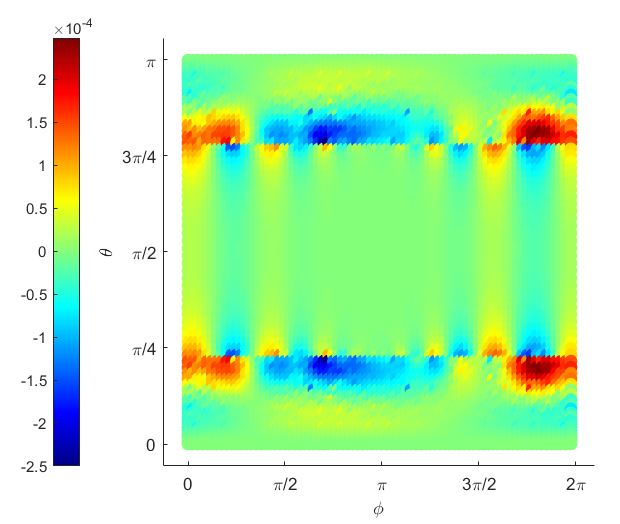}}}\hspace{5pt}
\subfloat[Imaginary]{%
\resizebox*{5.29cm}{!}{\includegraphics{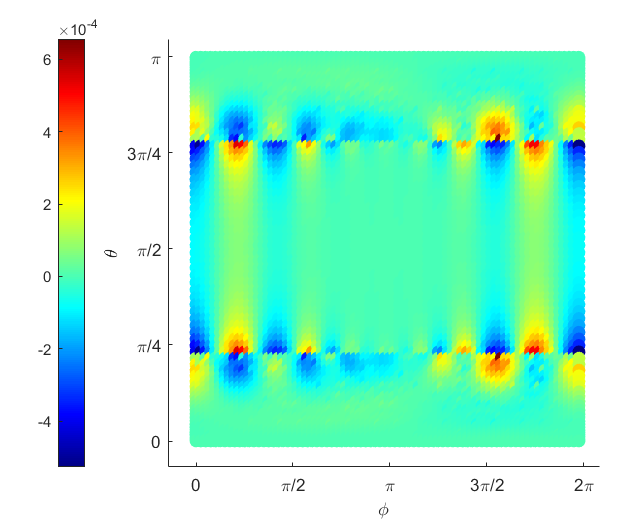}}}\hspace{5pt}
\caption{The computed normal velocity on the actual source} 
\label{ocean2_nv}
\end{figure}

The average radiated power of the source is around $1.8071 \times 10^{-5}$ or about 72.57 dB. Figure \ref{ocean2_nv} shows the corresponding normal velocity on the actual source. It can be observed that the maximum magnitude is just about $6.60 \times 10^{-4}$.

\subsubsection{A planewave in the near field}
In this experiment, we prescribe the plane wave $f(\mathbf x) = e^{i \mathbf x \cdot (10\mathbf d)}$ with $\mathbf d = \left [-1, 0, 0 \right ]$ on the near control. In the direction of $\hat \Bx_1$ we set a zero far field pattern value while in $\hat \Bx_2$ we prescribe a value of $0.05$. This mimics near field communication with minimal spill-over behind the near control while projecting a different far field signature in another direction.

\begin{figure}[!h]
\centering
\subfloat[Prescribed field]{%
\resizebox*{5.25cm}{!}{\includegraphics{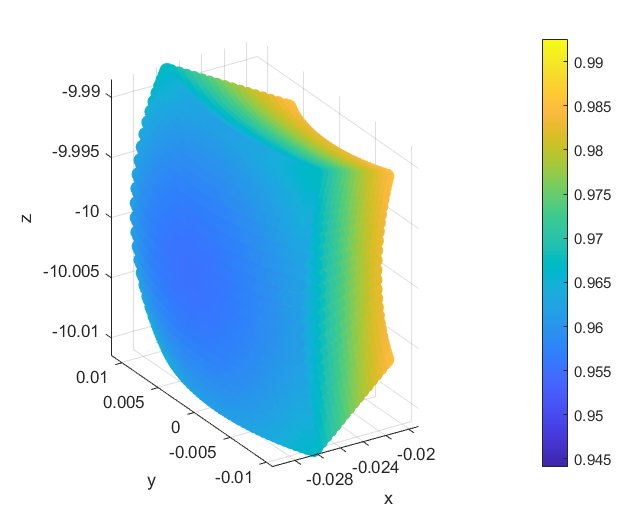}}}\hspace{5pt}
\subfloat[Generated field]{%
\resizebox*{5.25cm}{!}{\includegraphics{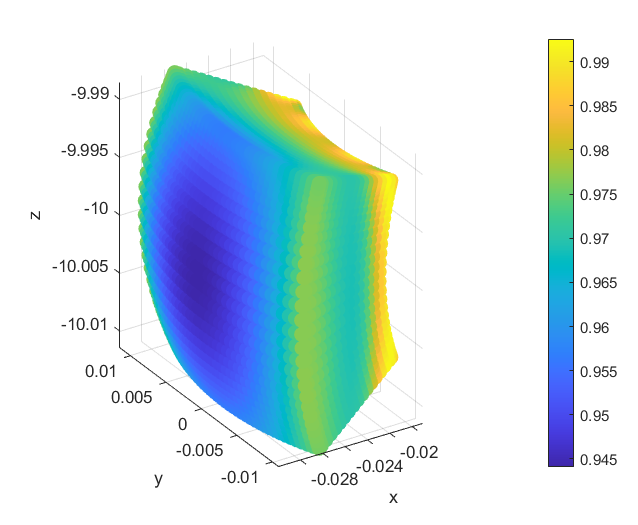}}}\hspace{5pt}
\subfloat[Relative error]{%
\resizebox*{5.25cm}{!}{\includegraphics{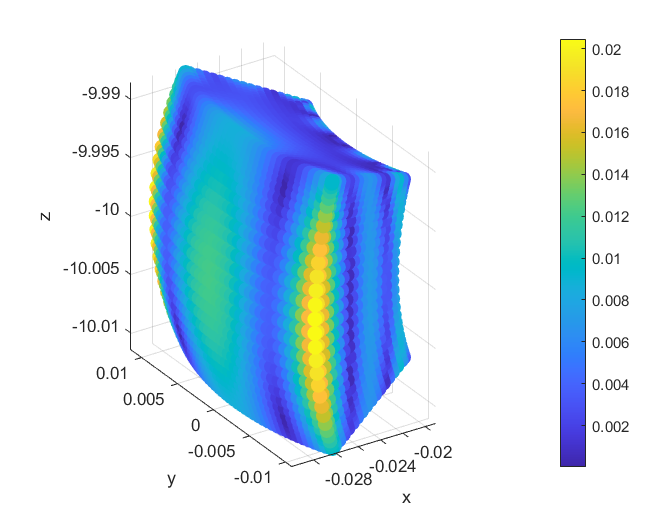}}}\hspace{5pt}
\caption{Results of the field synthesis on the near control} 
\label{ocean4_near}
\end{figure}

Figure \ref{ocean4_near} shows that the near field is approximated well with a pointwise relative error of at most 2.05\%.

\begin{figure}[!h]
\centering
\subfloat[Generated field on a patch around $\mathbf{\hat x_1}$]{%
\resizebox*{5.5cm}{!}{\includegraphics{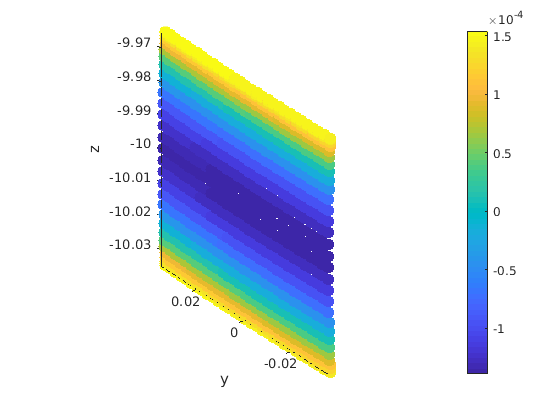}}}\hspace{5pt}\\ 
\subfloat[Generated field on a patch around $\mathbf{\hat x_2}$]{%
\resizebox*{5.5cm}{!}{\includegraphics{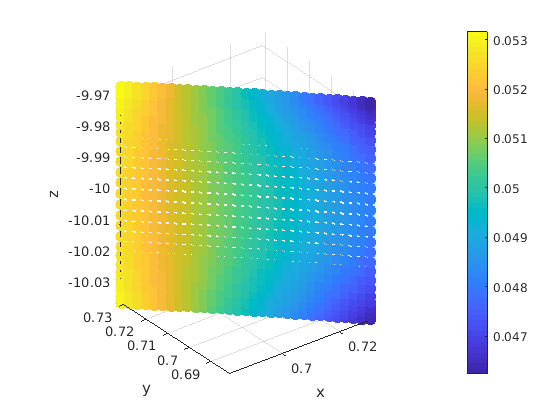}}}\hspace{5pt}
\subfloat[Relative difference from the prescribed value]{%
\resizebox*{5.5cm}{!}{\includegraphics{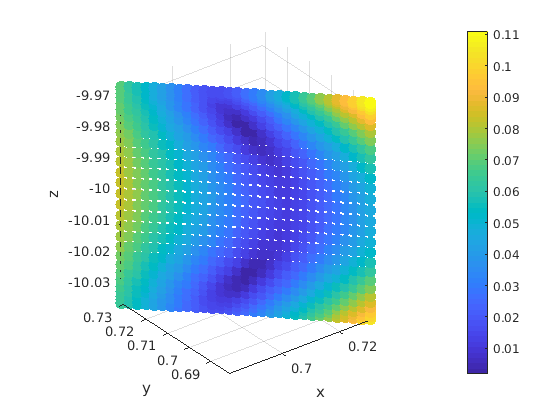}}}\hspace{5pt}
\caption{Results of the far field pattern directional control} 
\label{ocean4_far}
\end{figure}

Figure \ref{ocean4_far} shows the generated values on the patches around the directions $\hat \Bx_1$ and $\hat \Bx_2$. The values on the patch around $\hat \Bx_1$ are within order $10^{-4}$. In the exact direction $\hat \Bx_1$, the generated value has real part $-1.38 \times 10^{-4}$. Also, it can be noted that the relative errors on the patch around $\hat \Bx_2$ reach as high as 11\%. However, for points very near the exact direction $\hat \Bx_2$, the approximation becomes better. In fact at the exact direction, the generated value is $0.0503$ with a relative error of just about 0.60\%. 

\begin{figure}[!h]
\centering
\subfloat[Magnitude]{%
\resizebox*{5.29cm}{!}{\includegraphics{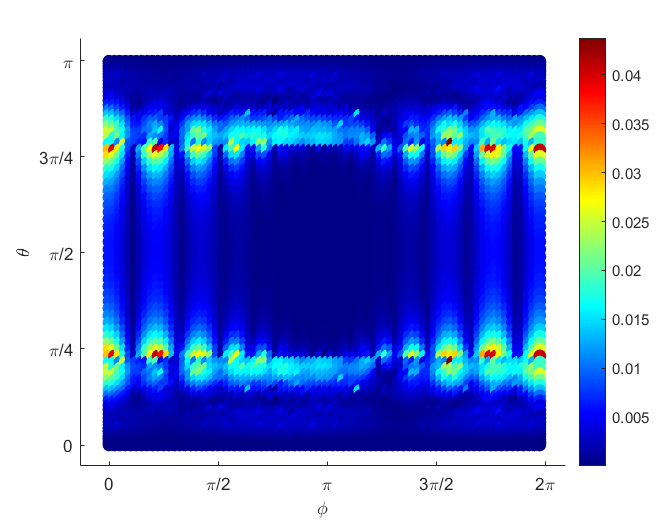}}}\hspace{5pt}
\subfloat[Real part]{%
\resizebox*{5.29cm}{!}{\includegraphics{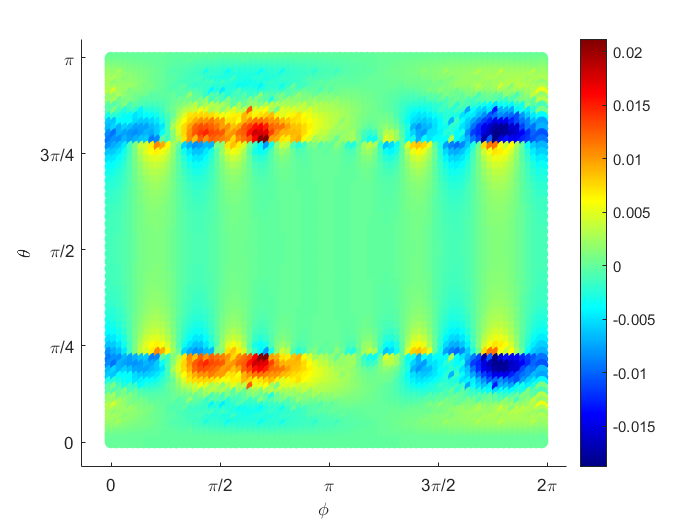}}}\hspace{5pt}
\subfloat[Imaginary]{%
\resizebox*{5.29cm}{!}{\includegraphics{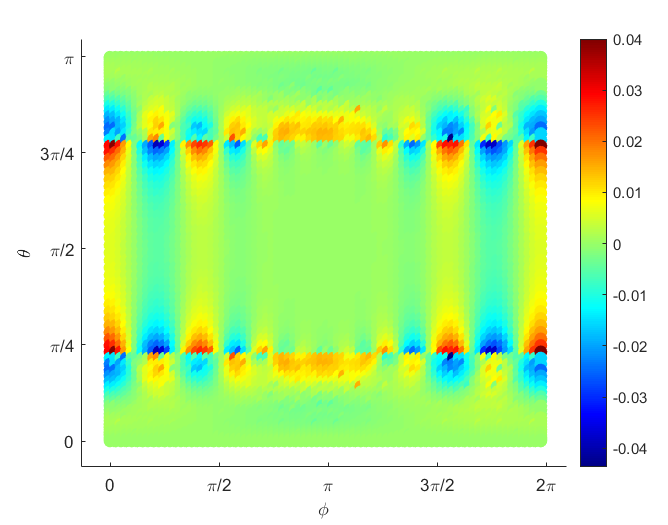}}}\hspace{5pt}
\caption{The computed normal velocity on the actual source} 
\label{ocean4_nv}
\end{figure}

The normal velocity on the physical source for this simulation is shown in Figure \ref{ocean4_nv}. The average radiated power by the source is around $9.97 \times 10^{-2}$ or about 109.99 dB.

\subsubsection{Two near controls and two far field directions}
\begin{figure}[!h]
\centering
\includegraphics[width=0.5\textwidth]{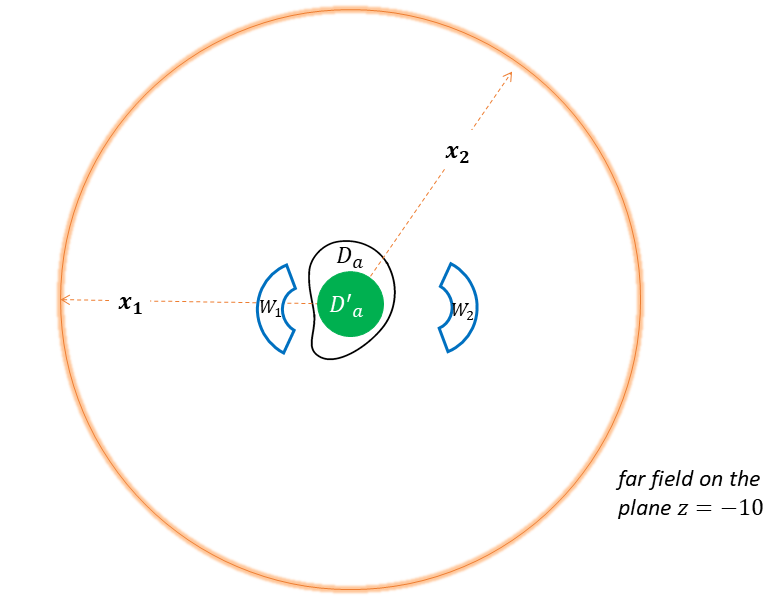}
\caption{A cross section of the simulation geometry}
\label{ocean5_geom}
\end{figure}

In this simulation, we consider an additional near control. Now, we have two near controls (given in spherical coordinates with respect to the source's center and where $\theta$ is the inclination while $\phi$ is the azimuthal angle) \[ W_1=\left \{(r, \theta, \phi): r \in [0.02, 0.03],  \theta \in \left [ \frac{\pi}{4}, \frac{3\pi}{4} \right ], \phi \in \left [ \frac{3 \pi}{4}, \frac{5 \pi}{4} \right ] \right \}\] and \[W_2 =\left \{(r, \phi, \theta) : r \in [0.15, 0.2],  \theta \in \left [\frac{\pi}{4}, \frac{3\pi}{4} \right ] , \phi \in \left [-\frac{\pi}{4}, \frac{\pi}{4} \right] \right \}.\]  The far field directions are still given by $\hat \Bx_1 =(1, \pi, -10)$ and $\hat \Bx_2 =\left (1, \frac{\pi}{4}, -10 \right )$ in cylindrical coordinates. A cross section of this problem geometry is shown in Figure \ref{ocean5_geom}. 

For this simulation we prescribe the outgoing planewave  $f(\mathbf x) = e^{i \mathbf x \cdot (10\mathbf d)}$ with $\mathbf d = \left [-1, 0, 0 \right ]$ on $W_1$ and a null field on $W_2$. Then at the direction $\hat \Bx_1$, we prescribe a zero far field pattern value and at $\hat \Bx_2$ we try to generate 0.05. This test simulates near field communication on $W_1$ with minimal spill-over in the direction behind it while keeping $W_2$ a quiet zone and projecting a decoy pattern in the far field direction $\hat \Bx_2$. 

\begin{figure}[!h]
\centering
\subfloat[Prescribed field]{%
\resizebox*{5.25cm}{!}{\includegraphics{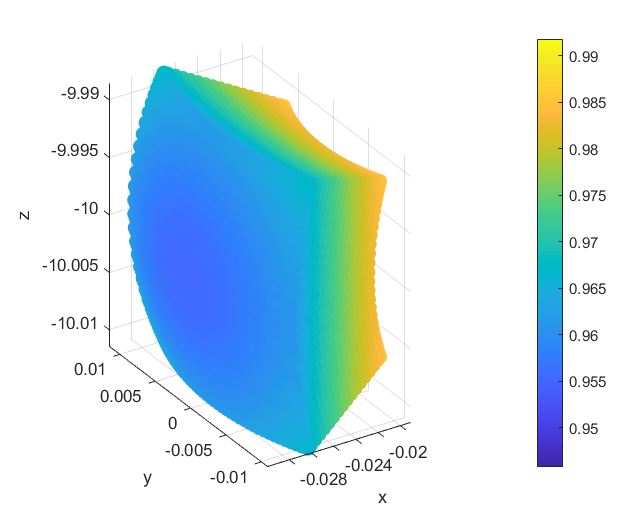}}}\hspace{5pt}
\subfloat[Generated field]{%
\resizebox*{5.25cm}{!}{\includegraphics{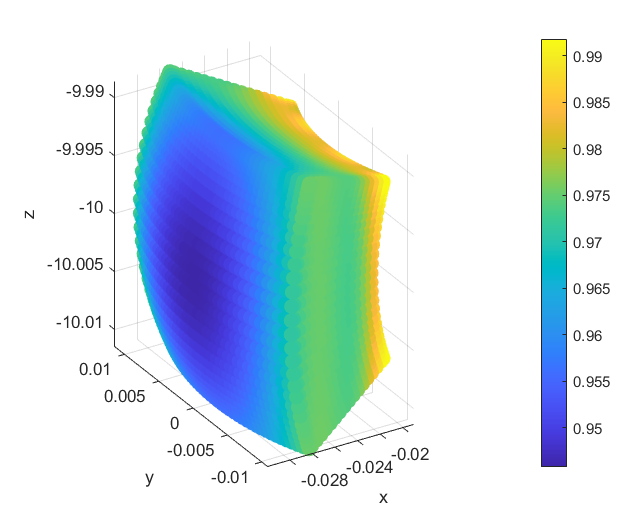}}}\hspace{5pt}
\subfloat[Relative error]{%
\resizebox*{5.25cm}{!}{\includegraphics{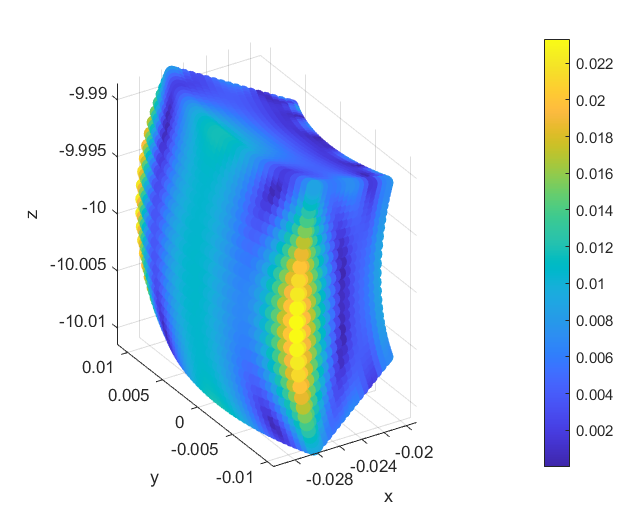}}}\hspace{5pt}
\caption{Results of the field synthesis on $W_1$} 
\label{ocean5_near}
\end{figure}

The results on $W_1$ are shown in Figure \ref{ocean5_near}. The first two plots show a visual comparison between the real parts of the prescribed and generated fields. The third plot shows the pointwise relative error. It can be observed that the relative errors are less than 2.33\% all throughout $W_1$. 

\begin{figure}[!h]
\centering
\includegraphics[width=0.4\textwidth]{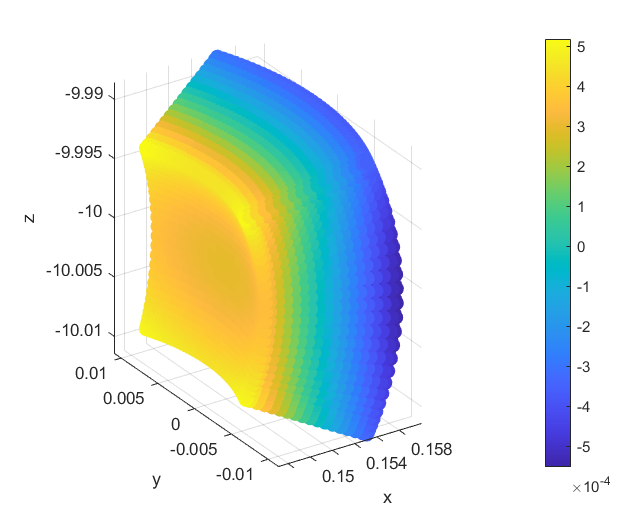}
\caption{Real part of the generated near field on $W_2$}
\label{ocean5_near2}
\end{figure}

Good results were likewise obtained for $W_2$. Figure \ref{ocean5_near2} shows that the generated field on the second near control is of order $10^{-4}$.

\begin{figure}[!h]
\centering
\subfloat[Generated field on a patch around $\mathbf{\hat x_1}$]{%
\resizebox*{5.5cm}{!}{\includegraphics{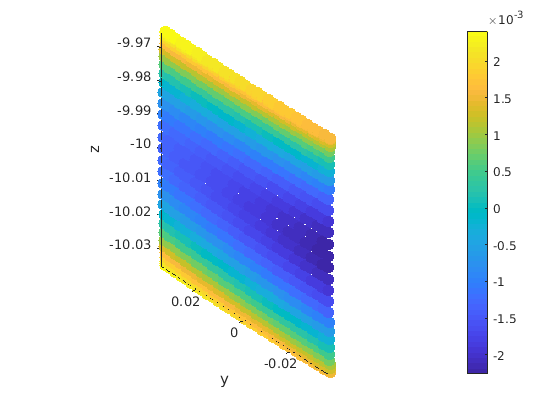}}}\hspace{5pt}\\ 
\subfloat[Generated field on a patch around $\mathbf{\hat x_2}$]{%
\resizebox*{5.5cm}{!}{\includegraphics{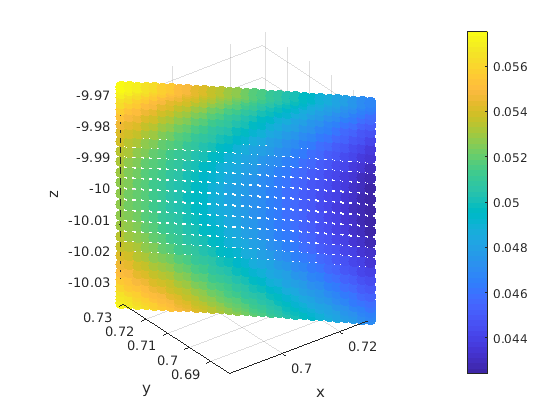}}}\hspace{5pt}
\subfloat[Relative difference from the prescribed value]{%
\resizebox*{5.5cm}{!}{\includegraphics{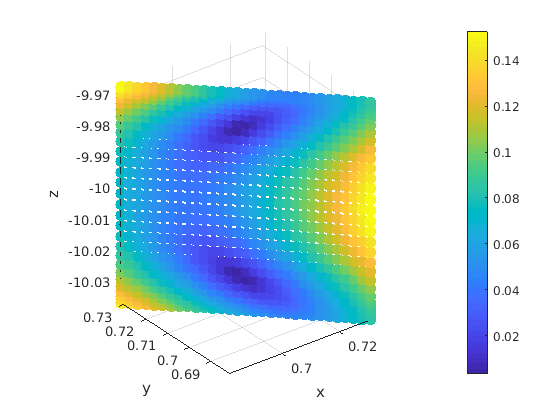}}}\hspace{5pt}
\caption{Results of the far field pattern directional control} 
\label{ocean5_far}
\end{figure}

In Figure \ref{ocean5_far}, the generated far field pattern values on small patches around the directions $\hat \Bx_1$ and $\hat \Bx_2$ are shown. The values around $\hat \Bx_1$ are all of order $10^{-3}$. At the exact direction, the generated value is an order smaller at $4.3 \times 10^{-4}$. The decoy pattern is matched well in a smaller subset of the patch around $\hat \Bx_2$. Nevertheless, in the exact direction, the generated value is $0.05061$ with relative error of only $1.22\%$.

\begin{figure}[!h]
\centering
\subfloat[Magnitude]{%
\resizebox*{5.29cm}{!}{\includegraphics{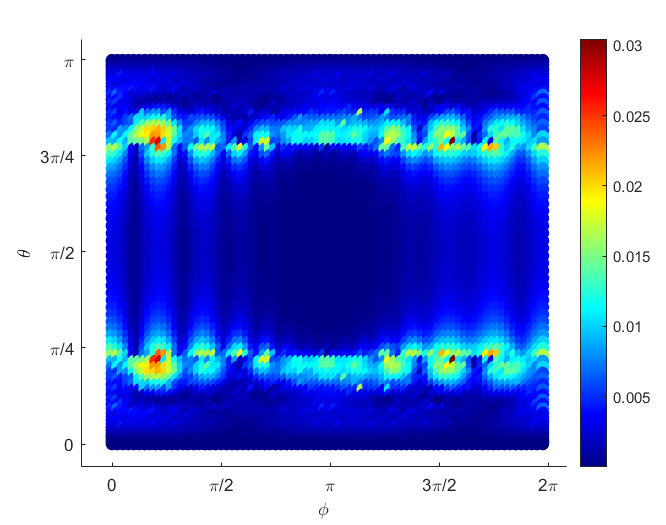}}}\hspace{5pt}
\subfloat[Real part]{%
\resizebox*{5.29cm}{!}{\includegraphics{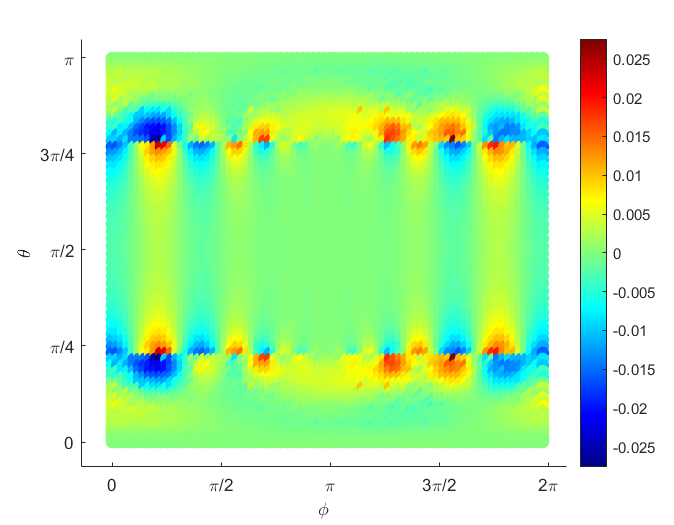}}}\hspace{5pt}
\subfloat[Imaginary]{%
\resizebox*{5.29cm}{!}{\includegraphics{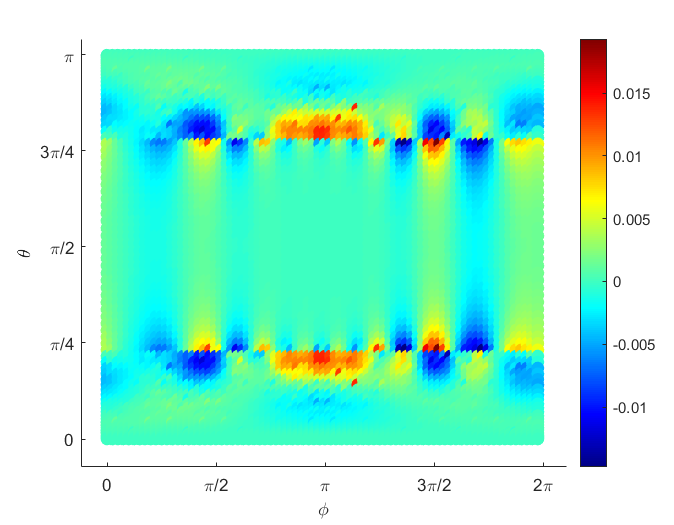}}}\hspace{5pt}
\caption{The computed normal velocity on the actual source} 
\label{ocean5_nv}
\end{figure}

The computed normal velocity on the actual source is described in Figure \ref{ocean5_nv}.  The average power radiated by the source is about $3.62 \times 10^{-2}$ or roughly 105.58 dB, a bit lower than the one obtained in the previous simulation. 

\section{Conclusion and Future Works}
\label{conclusions}

In this paper, we extended the theoretical results and the numerical schemes developed in our previous works on the active control of acoustic fields. We proved the possibility of controlling the acoustic field in the near field of an active source while doing a far field pattern control in multiple directions in both the free space and a homogeneous finite-depth ocean environment. This was done by showing that for any set of prescribed fields in multiple bounded control regions in the near field and prescribed far field patterns in distinct directions, one can always find a boundary input on the source, for instance the acoustic pressure on the surface of the source, that will approximate these prescribed fields.

Several numerical simulations in both environments were presented to illustrate the feasibility of the proposed framework. These simulations mimic scenarios in the development of enhanced communication strategies with focus on signal protection and interference avoidance. The results show a good approximation of the desired effects. In all these tests, the source seems to radiate a low average acoustic power. 

Our current numerical tests suggest that the solution is stable with respect to various geometric parameters as long as these parameters are within certain problem dependent ranges. In a forthcoming article, we shall provide a sensitivity analysis of our scheme with respect to variations in the frequency and changes in the problem geometry such as the size of the control regions, their distances from the source as well as the number of far field directions and regions of control and their relative positions. Another future research direction is the use of an array of coupling sources (with fixed or optimized locations) instead of one single source to mitigate possible high amplitudes needed on the boundary input on a single source.  The authors are also working on the extension of the results presented for the homogeneous ocean environment to a multi-layered ocean environment. A feasibility study on the possibility of physically instantiating the boundary inputs computed using the strategy proposed here is also forthcoming. These research directions may be aligned with interesting applications such as enhanced communications in free space and underwater environments.

\section{Acknowledgments}
D. Onofrei and N. J. A. Egarguin would like to acknowledge the Army Research Office, USA for funding their work under the award W911NF- 17-1-0478. J. Chen and C. Qi would like to acknowledge the National Science Foundation, USA for funding their work under the award 1801925.

\bibliographystyle{plain}
\bibliography{fardirpaper}


\end {document}